\documentclass[manuscript]{acmart}

\usepackage{makecell}
\usepackage{url}
\usepackage[caption=false,font=footnotesize]{subfig}
\DeclareUnicodeCharacter{2212}{-}

\newcommand{\KBS}{KnowledgeShovel}

\newcommand{\SubItem}[1]{
    {\setlength\itemindent{15pt} \item[-] #1}
}

\AtBeginDocument{%
  \providecommand\BibTeX{{%
    \normalfont B\kern-0.5em{\scshape i\kern-0.25em b}\kern-0.8em\TeX}}}

\begin{document}

\title{\KBS: An AI-in-the-Loop Document Annotation System for Scientific Knowledge Base Construction}

\author{Shao Zhang}
\email{shaozhang@sjtu.edu.cn}
\affiliation{
  \institution{Shanghai Jiao Tong University}
  \city{Shanghai}
  \country{China}
}

\author{Yuting Jia}
\email{hnxxjyt@sjtu.edu.cn}
\affiliation{
  \institution{Shanghai Jiao Tong University}
  \city{Shanghai}
  \country{China}
}

\author{Hui Xu}
\email{xhui_1@sjtu.edu.cn}
\affiliation{
  \institution{Shanghai Jiao Tong University}
  \city{Shanghai}
  \country{China}
}

\author{Dakuo Wang}
\email{dakuo.wang@ibm.com}
\affiliation{
  \institution{IBM Research}
  \city{Cambridge, Massachusetts}
  \country{United States}
}

\author{Toby Jia-Jun Li}
\email{toby.j.li@nd.edu}
\affiliation{
  \institution{University of Notre Dame}
  \city{Notre Dame, Indiana}
  \country{United States}
}
\author{Ying Wen}
\email{ying.wen@sjtu.edu.cn}
\affiliation{
  \institution{Shanghai Jiao Tong University}
  \city{Shanghai}
  \country{China}
}
\author{Xinbing Wang}
\email{xwang8@sjtu.edu.cn}
\affiliation{
  \institution{Shanghai Jiao Tong University}
  \city{Shanghai}
  \country{China}
}

\author{Chenghu Zhou}
\email{zhouch@lreis.ac.cn}
\affiliation{
  \institution{Chinese Academy of Sciences}
  \city{Beijing}
  \country{China}
}

\begin{teaserfigure}
    \centering
    \includegraphics[width=0.8\linewidth]{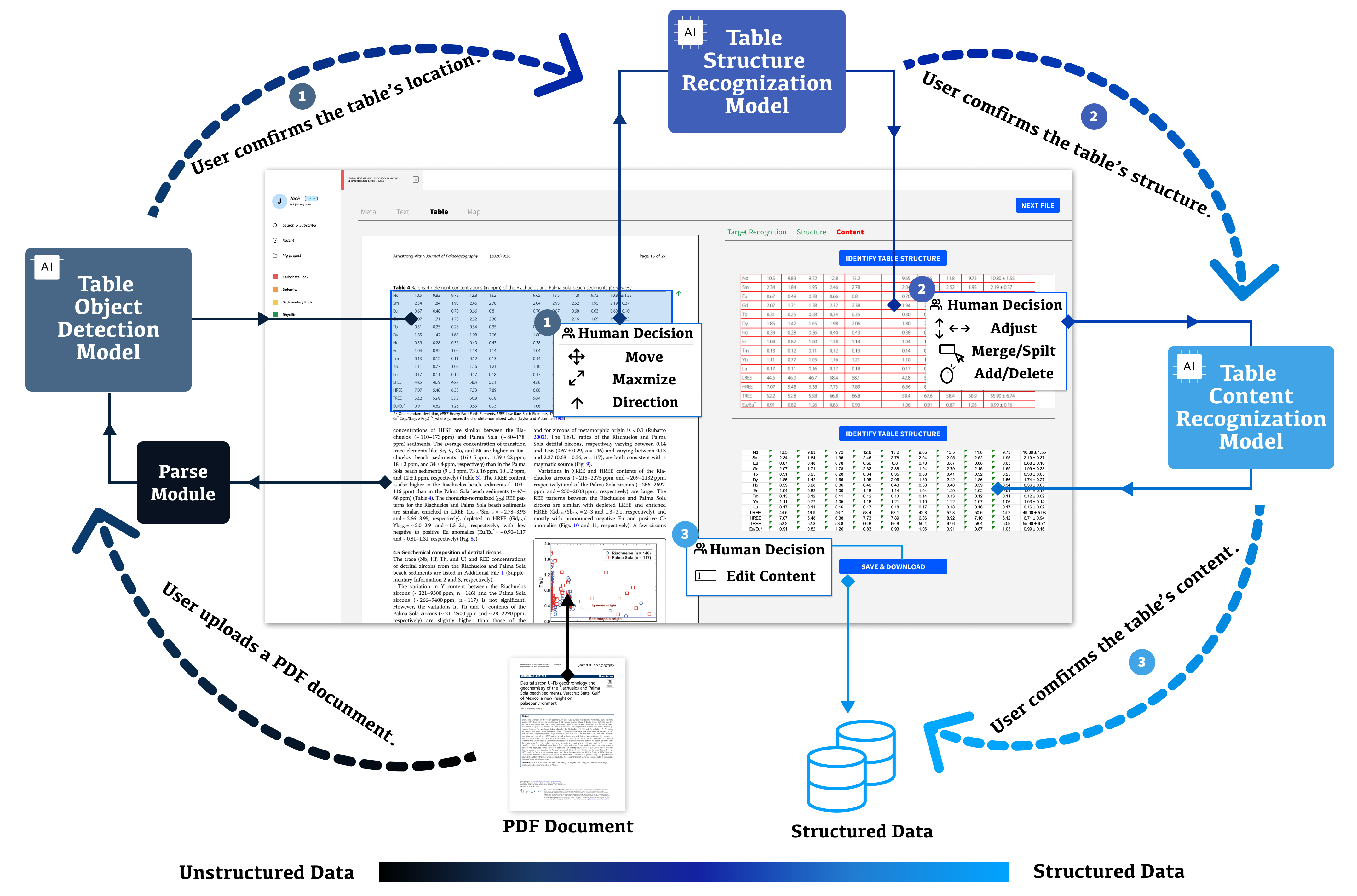}
    \caption{\textbf{The AI-in-the-loop thinking of multi-step human-AI collaboration pipeline in \KBS~to obtain structured knowledge base from unstructured data.} The pipeline starts with the PDF Document, and it involves three AI modules for this table extraction feature demonstration. \KBS is designed to visualize the AI's intermediate output to humans so that humans may verify and correct the results, which effectively prevents the accumulation of errors in this multi-stage multi-modality pipeline, and the final knowledge extraction can meet users' high precision requirement.  }
    \label{fig:table_extraction_interaction}
    \Description{The figure shows the table extraction process in KnowledgeShovel. The pipeline uses a form that effectively interprets the model's output to humans that can push humans to correct and validate the results. The human-AI collaboration pipeline has the following 6 steps: (1) AI model pre-located the position of each table in the PDF; (2) User chooses a table and confirm the table area; (3) AI model recognizes the table's structure; (4) User gets the table's structure result from AI model and confirm; (5) AI model recognizes the table's content; (6) User edits and confirms the content. In the process, the extracted data becomes more and more structed.}
\end{teaserfigure}

\renewcommand{\shortauthors}{Zhang et al.}

\begin{abstract}
Constructing a comprehensive, accurate, and useful scientific knowledge base is crucial for human researchers synthesizing scientific knowledge and for enabling AI-driven scientific discovery. 
However, the current process is difficult, error-prone, and laborious due to (1) the enormous amount of scientific literature available; (2) the highly-specialized scientific domains; (3) the diverse modalities of information (text, figure, table); and, (4) the silos of scientific knowledge in different publications with inconsistent formats and structures. 
Informed by a formative study and iterated with participatory design workshops, we designed and developed \textbf{\KBS}, an AI-in-the-Loop document annotation system for researchers to construct scientific knowledge bases. 
The design of \KBS~introduces a multi-step multi-modal human-AI collaboration pipeline that aligns with users' existing workflows to improve data accuracy while reducing the human burden. 
A follow-up user evaluation with 7 geoscience researchers shows that \KBS~can enable efficient construction of scientific knowledge bases with satisfactory accuracy.

\end{abstract}

\ccsdesc[500]{Human-centered computing~Human computer interaction (HCI)}
\ccsdesc[500]{Applied computing~Annotation}
\ccsdesc[300]{Computing methodologies~Information extraction}
\keywords{AI-in-the-Loop, Human-AI Collaboration, Data Extraction, Scientific Knowledge Base, Scientific Literature Processing}

\maketitle

\section{Introduction}

Scientific research is currently advancing faster than ever before, with tens of thousands of scientific studies published every day in various forms such as papers, preprints, and datasets.
\textbf{Scientific knowledge bases} \cite{national2000question,hoeppe2021encoding}, a collection of structured and verified research results that consists of various numeric, word-oriented, or image-organized data, emerge in this context and bring entirely new approaches and opportunities to scientific research.
Researchers in many disciplines uses AI techniques and the scientific knowledge bases, often constructed from the published literature, to drive scientific discoveries \cite{PUETZ2018877,mcmahon2018evolution,puetz2018statistical}, such as Geoscience \cite{bergen2019machine,10.1145/3060586}, Medicine \cite{austin2016application}, Biology \cite{altaf2014systems},  Chemistry \cite{schleder2019ab}.
The rapid development of AI and data science has further promoted the development of scientific knowledge base \cite{hope-etal-2021-extracting,murthy2022accord}.
For example, AlphaFold \cite{jumper2021highly}, which uses Protein Data Bank \cite{10.1093/nar/gky949} as input data, can accurately predict protein structure and greatly promote the development of biological and medical research \cite{mirdita2022colabfold,bryant2022improved}.
\begin{figure}
    \centering
    \includegraphics[width=\linewidth]{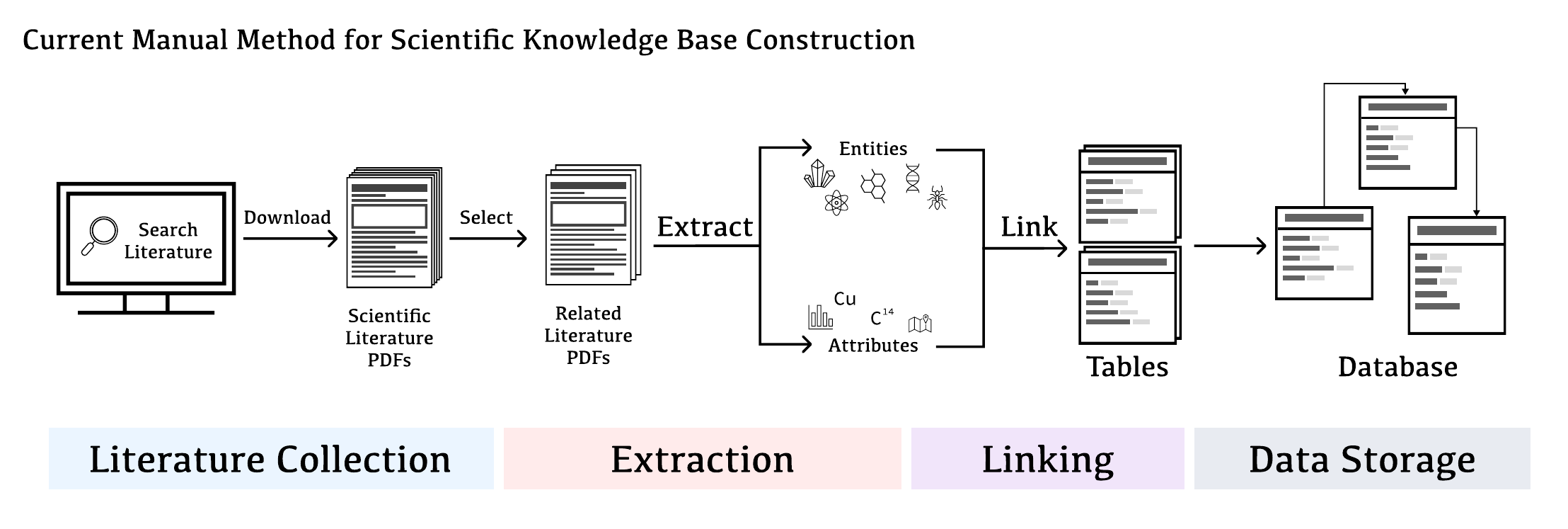}
    \caption{The current practice of domain experts manually constructing a scientific knowledge base.}
    \label{fig:manual}
\end{figure}

Although successful research examples illustrate the importance of scientific knowledge bases for scientific research in the data explosive age, there are still many challenges in the composition of the scientific knowledge base and the construction process due to their characteristics. 
The characteristic of a scientific knowledge base composition is that it is described around one type of scientific entity.
For example, ``sample'' is a general type of scientific entity. 
The data contained are the values and sources of the relevant attributes of the scientific entity.
The current process of constructing a scientific knowledge base includes four main steps:\textit{literature collection}, \textit{entity and attribute extraction}, \textit{entity linking}, and \textit{data storage} (see Figure \ref{fig:manual}).
The task of building a scientific knowledge base has the following four characteristics:
(1) \textbf{Needs of an enormous amount of scientific literature}: In order to cover more scientific entities to describe in depth and breadth, a scientific knowledge base needs a large number of source documents for construction, which means an incredible workload and data processing difficulty;
(2) \textbf{Under a highly-specialized scientific domain}: Due to constructing in highly-specialized scientific domains, a scientific knowledge base can only be constructed by domain experts in selecting the appropriate sources, extracting and linking the specific entities;
(3) \textbf{Diverse modalities of information}: Scientific articles generally use PDF as a standard format, which contains data of multiple modalities such as text, tables, pictures, etc., and leads to a complicated extraction and linking process;
(4) \textbf{Silos of scientific knowledge}: The scientific knowledge in different publications with inconsistent formats and structures \cite{10.1145/2851581.2892588} leads to very low efficiency of manual extraction. 
Moreover, a scientific knowledge base that complies with the FAIR principle (findable, accessible, interoperable, and reusable) \cite{wilkinson2016fair} can promote the openness of data and the reuse of data in disciplinary research, which also challenges the reusability and simplicity of the knowledge base construction approach.

\begin{figure}
\centering
\subfloat[Current Human-in-the-Loop Method for Scientific Discovery]{\includegraphics[width=\linewidth]{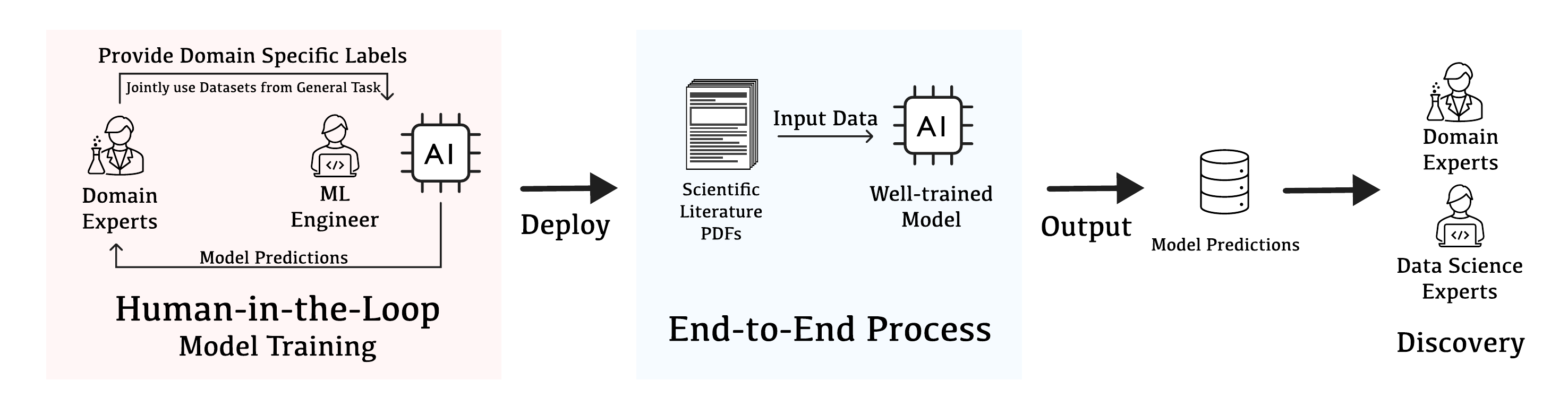}
\label{fig:hitl}}

\subfloat[\KBS’s AI-in-the-Loop Method of Building Scientific Knowledge Base for Scientific Discovery]{\includegraphics[width=\linewidth]{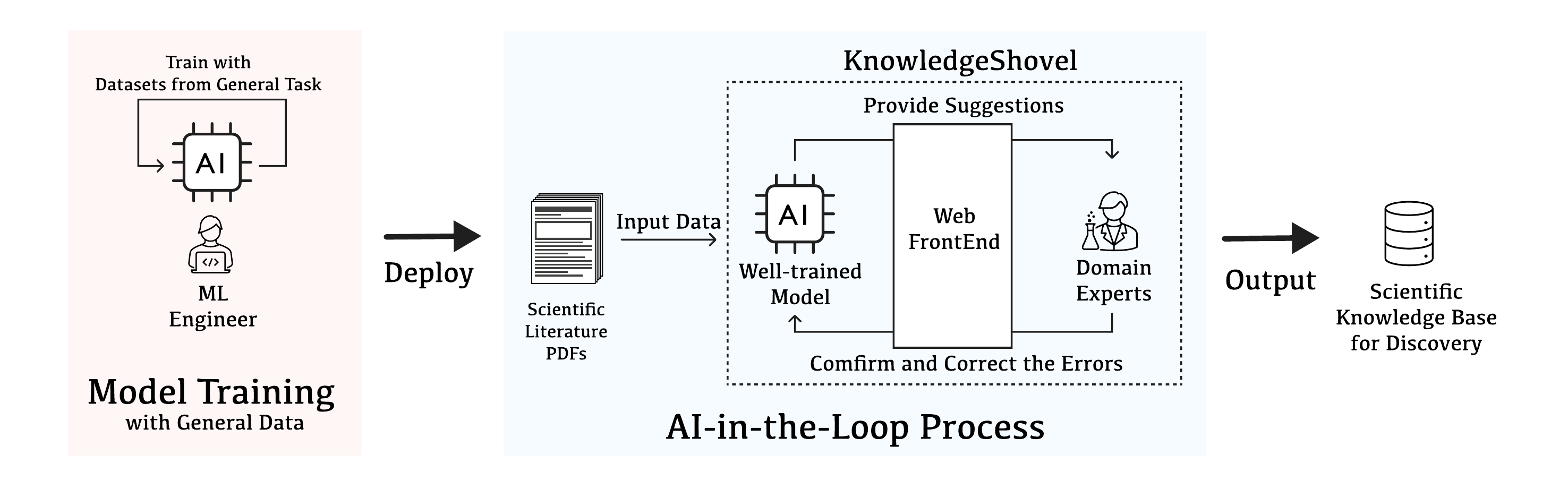}
\label{fig:aiitl}}
\caption{Human-in-the-loop VS AI-in-the-loop. Human-in-the-loop is a method for improving model performance, which requires humans to provide labeled data, adjust the final outputs of the model, and make greater efforts to ensure the accuracy of the data. AI-in-the-loop is a method where AI is input and reflected into human systems and helps humans to build a scientific knowledge base, which can reduce the workload of humans in the construction of scientific knowledge bases requiring extremely high data accuracy.}
\label{fig:hitl-vs-aiitl}
\end{figure}

Researchers have explored the human-in-the-loop paradigm (see Figure \ref{fig:hitl}) for data extraction tasks, and some efforts have been practiced in the scientific knowledge bases construction use case  \cite{wu2022survey,ristoski2020large} to reduce the burden on researchers and improve the efficiency.
Extracting and processing the unstructured ``data'' in the scientific literature and transferring it to structured data (from entity and attribute extraction to entity linking) is the most challenging step.
Thus, the human-in-the-loop approach is designed to address the difficulties in this step.
However, the human-in-the-loop AI architecture still cannot meet the requirements of scientific knowledge bases construction, which is mainly due to \textbf{the limitations of AI capabilities}, \textbf{difficulty in obtaining labeled data}, and \textbf{the non-interoperability of models}.

In this paper, we aim to explore a novel approach that an AI-based solution can overcome the aforementioned limitations and can provide desired helps to support real-world scientists in constructing a scientific knowledge base.
We started our effort with a formative study with 12 participants from 9 geoscience groups as a case study.
Driven by the formative study findings, we arugue it is essential to explore the AI-in-the-loop method (refers to Figure \ref{fig:aiitl}) to help researchers construct scientific knowledge bases.
In the AI-in-the-loop approach, the workflow is derived from the human workflow, thus domain experts can directly interact with the model at their convenient times, so they have a reasonable mental model of how the AI capabilities should work.
Based on the formative study findings, we proposed various design guidelines, and then iterated them through a participatory design session with 14 participants.

Building on prior literature, our formative study, and participatory design sessions with 9 groups of researchers, we designed and implemented \textbf{\KBS}, an AI-in-the-loop system that helps researchers intuitively annotate literature and extract data to construct a scientific knowledge base. 
The system consists of a web front-end that follows the guidelines of human-AI interaction \cite{10.1145/3290605.3300233}, a back-end server, and some basic models for data extraction. 
We proposed a two-level data processing workflow with the multi-step human-AI collaboration pipeline using the web front-end (see Figure \ref{fig:table_extraction_interaction}), which can balance efficiency and data accuracy, and help build a better mental model of AI for humans.
The first level is the single-document-level data extraction, where researchers can interact with AI to get assistance for extracting data including meta information, tables, texts, and figures.
The next level is the project-level data integration, where the contents of single-document-level data extraction are fused to form a knowledge base.
We deployed and granted \KBS~ access to researchers in the geoscience departments from more than 10 universities, then conducted a formal user study with 7 experts.
The user study results show that the AI-in-the-loop method can successfully improve the efficiency and gain a highly accurate data.

This paper makes the following contributions:
\begin{itemize}
    \item a formative study and a participatory design process with researchers working on building scientific knowledge bases that reveals the current workflow of researchers in professional fields, the challenges they face, and their needs for data accuracy and extraction efficiency in the process of building scientific knowledge bases.
    \item the design and implementation of \KBS, an AI-in-the-loop system where researchers collaborate with AI in extracting data efficiently from literature to build a scientific knowledge base with high data quality without the dependence of data scientists .
    \item a user study with 7 domain researchers that evaluates the usability of \KBS~and sheds light on how researchers use \KBS~to improve their work and how AI-in-the-loop influence the scientific knowledge base construction.
\end{itemize}

\section{Background and Related Work}
\subsection{What is a Scientific Knowledge Base?}

Different from the concept of the database in computer science, a scientific knowledge base is a collection of data generated in the process of scientific research \cite{1701919}. 
Scientific knowledge bases are now widely used in big data-driven research in many disciplines, including Geoscience \cite{bergen2019machine}, Medicine \cite{austin2016application}, Biology \cite{altaf2014systems},  Chemistry \cite{schleder2019ab}.
A scientific knowledge base is usually described around a scientific entity and mainly contains the relevant attribute values of this entity. 
For example, in geosciences, \textit{``A relational database of global U–Pb ages''} \cite{PUETZ2018877} is an example of one kind of scientific knowledge base that contains 700,598 records of global U-Pb ages with a main scientific entity \textit{``sample''}.
It is also a scientific knowledge base constructed using data extracted from the literature \cite{PUETZ2018877}.  
Similarly, \cite{mcmahon2018evolution} and \cite{puetz2018statistical} are the research using scientific knowledge bases constructed by data extracted from the literature.

As big data-driven research becomes commonplace, an increasing amount of scientific knowledge bases are needed, and building a reusable database becomes a challenge for researchers \cite{wilkinson2016fair}.
Researchers have pointed out that the process of data preparation is vital throughout the research process \cite{sun2022review}.
Furthermore, many researchers think building a system for data extraction access is necessary for a scientific knowledge base \cite{davydov2006detail}.

In this paper, we aim to reveal empirical understandings and develop an AI-in-the-loop system to support researchers' data extraction work for building a scientific knowledge base.
Specifically, we investigated 1) a formative study to understand how researchers extract data from scientific literature,
2) to implement a system with AI assistance to support their work, and 3) to explore researchers' feedback and design implications from the user study.

\subsection{Scientific Data Extraction from Literature}

As we discussed before, the construction of scientific knowledge bases has important implications for research. 
Many current disciplines spend much time extracting data, but the time and effort spent in these processes are often invisible.

Artificial intelligence methods are introduced in the data extraction from literature to help promote the data extraction from the published literature and build a scientific knowledge base in Biology \cite{li2014biological}, Chemistry \cite{swain2016chemdataextractor,rajan2021decimer}, and many other disciplines \cite{hong2021challenges,walker2022evaluation,decision-making-medical}.
However, most data extraction studies using artificial intelligence in the scientific literature require case customization. 
Customized cases can only solve a single knowledge base construction task, which requires the in-depth participation of artificial intelligence experts, and it is not easy to expand to a broader range of research.

The more general solution and tookits only stays in the scientific literature parsing process, including Grobid \cite{GROBID}, Science Parse \cite{tkaczyk2018machine} and PDFFigures 2.0 \cite{clark2016pdffigures}.
We find in formative research that domain researchers know very little about AI techniques, which hinders these methods from working.
The lack of graphical interfaces makes it difficult for researchers who are only average computer users to use these tools and efficiently carry out further data processing \cite{lanius2021usability}.
Although some existing tools can help with extracting the table and figures in PDF documents with graphical interfaces \cite{wang2021tablelab,finereaderpdf}, without a design for scientific literature, these tools are not suitable for processing scientific literature PDFs and do not accomplish the extraction and linking of data.

The current data extraction methods used for literature retrieval and dissertation comprehension \cite{luan2018information,tkaczyk2015cermine} can be used for reference in scientific knowledge base construction,  but there are still challenges to solve.
For example,  Google Scholar \cite{googlescholar}, Semantic Scholar \cite{ammar2018construction}, and AceMap \cite{tan2016acemap} are built based on literature, using natural language processing methods to classify the documents and extract features by parsing the documents to achieve the function of searching and recommending documents.
Literature processing in such a method is limited to the text part without the pictures and tables' data.
The other problem is that the main object of the database is the article itself rather than a scientific entity due to the functionality consideration, which means that, like the other parsing tools, these methods do not drill down to the scientific data contained within the literature.
The academic search system database contains only the article's attribute information, such as the journal author abstract, etc., which can only be used for model training related to search and recommendation \cite{ammar2018construction}, but not possible to be used to build a scientific knowledge base.

\subsection{Human-in-the-loop Data Extraction}

Many studies have attempted to use a human-in-the-loop modeling training approach with end-to-end extraction to enhance accuracy and reduce the human burden.
There are currently many machine learning systems that use a human-in-the-loop approach to involve people in the training and deployment process of the model and use human feedback to label data to improve the performance of the model \cite{wu2022survey}.
For the data extraction task, there are also many human-in-the-loop methods to improve the performance of the model from collecting human feedback \cite{10.1145/3404835.3463247,ristoski2020large}, collecting labeled data \cite{zhang2019invest}, and interactive training \cite{10.1145/3184558.3191546}.
However, we can find that these human-in-the-loop methods only evaluate the final artificial intelligence model's performance but ignore the human experience in this process and the application's performance in the final deployment stage.

In the HCI community, research related to human-in-the-loop and human-AI collaboration has explored issues such as the burden of humans in the process of labeling tasks \cite{do-model-work} and the experience of crowdworkers \cite{wang2022whose}.
There are also some studies on efficiency improvement and factors affecting the labeling process \cite{ashktorab2021ai}.
However, all the studies above are aimed at the model training process. 
In the model deployment stage, during the process of human interaction with the model, the human-in-the-loop process often faces problems that are different from the training process.
We found in formative research that many researchers do not benefit from the human-in-the-loop approach.
Researchers spend similar manual extraction efforts to annotate the data to support the human-in-the-loop approach during the model training phase, but cannot obtain the high-precision data needed during the deployment phase.
They still need to spend more time proofreading the data output by the model to ensure the accuracy of the data.
The main reason for this is that the gap between the amount of data required to train the model and the amount of data that needs to be processed in practical applications is too small. 
In other words, the data required to train a high-precision model in a human-in-the-loop manner is equivalent to the amount of data that the model needs to be deployed to process.

Currently, in some real-time systems, there are already relevant practices on AI-in-the-loop \cite{gheibi2021helping,cholia2021integrating}, and we think this provides a reference for introducing human-AI collaboration in the construction of scientific knowledge bases.
To solve this problem, \KBS~uses the AI-in-the-loop approach to model deployment and application. 
We no longer require experts to spend much time on human-in-the-loop data labeling to improve the model's accuracy, but use a general model to assist experts in the direct extraction of data.

\section{Formative Study}
We aim to design a novel approach to support researchers construct a scientific knowledge base, so our first step is to explore the current practices and requirements.
We conducted a formative study following the methodology in~\cite{olson2014ways}. 
We first started with a questionnaire to investigate the relevant technical background of the domain experts and the data extraction practices involved in their daily work. 
The questionnaire was sent to 18 geoscience research groups and in total 119 (64 male)  people responded to our questionaire, of which 106 were valid (57 male). We further recruited  12 people for a 60-minutes in-depth interviews.

\subsection{Questionnaire Survey}\label{Q}

We choose to use questionnaire surveys \cite{muller2014survey} 
to understand the current status of work in the field of earth sciences on the task of building subject-specific databases.
We divided the questionnaire into three main parts (Appendix \ref{questinnaire}):
    basic information,
    task-related information,
    and, user's understanding of computer technology.

\subsection{In-depth Interview}\label{I}

Based on the questionnaire results, we designed the semi-structured interview questions, and recruited potential interviewees to conduct face-to-face interviews \cite{longhurst2003semi}. 
We organized the semi-structured interview with a schema of questions about their research interests, teamwork, and usage of the data they collected to explore their needs during the workflows.
We invited 12 users from 9 different research groups for 60 minutes interviews (each group) from the questionnaire survey participants who have different roles in their research groups.
The interviewees' demographic info and project experience are shown in Table \ref{tab:Interviewees}.
\begin{table}
    \centering
    \resizebox{1. \linewidth}{!}{
    \begin{tabular}{c|c|c|l|l}
    \toprule
    Group ID &  Participant ID & Gender & Research Projects & Role\\
    \midrule
      GA1   &  PA01 & Male & Magmatic Migration & PhD Student\\
      GA2   &  PA02 & Male & Geomagnetism and Geoelectromagnetism & Associate Professor\\
      GA2   &  PA03 & Male & Geomagnetism and Geoelectromagnetism & PhD Student\\
      GA3   &  PA04 & Male & Paleoclimatology & Associate Professor\\
      GA4   &  PA05 & Male & Geochronology and Structural Geology & Professor\\
      GA5   &  PA06 & Female & Paleontology & Associate Researcher\\
      GA6   &  PA07 & Male & Structural Geology & PhD Student\\
      GA7   &  PA08 & Female & Evolutionary Biology and Dinosauria & PhD Student\\
      GA7   &  PA09 & Male & Evolutionary Biology and Dinosauria & PhD Student\\
      GA8   &  PA10 & Male & Carbonate Sedimentology & Postdoctoral Researcher\\
      GA9   &  PA11 & Female & Global Detrital Zircon Database & Full-time Data Entry Assistant\\
      GA9   &  PA12 & Male & Global Detrital Zircon Database & Full-time Data Entry Assistant\\
  \bottomrule
    \end{tabular}
    }
    \caption{Demographics of interviewees.}
    \label{tab:Interviewees}
\end{table}

\subsection{Key Insights}
Through an open-coding process conducted by two co-authors in this team, we identified the following key insights of the geoscientists' existing workflow and their requirements for the scientific knowledge base construction.

\begin{figure}
    \centering
    \includegraphics[width=0.95\linewidth]{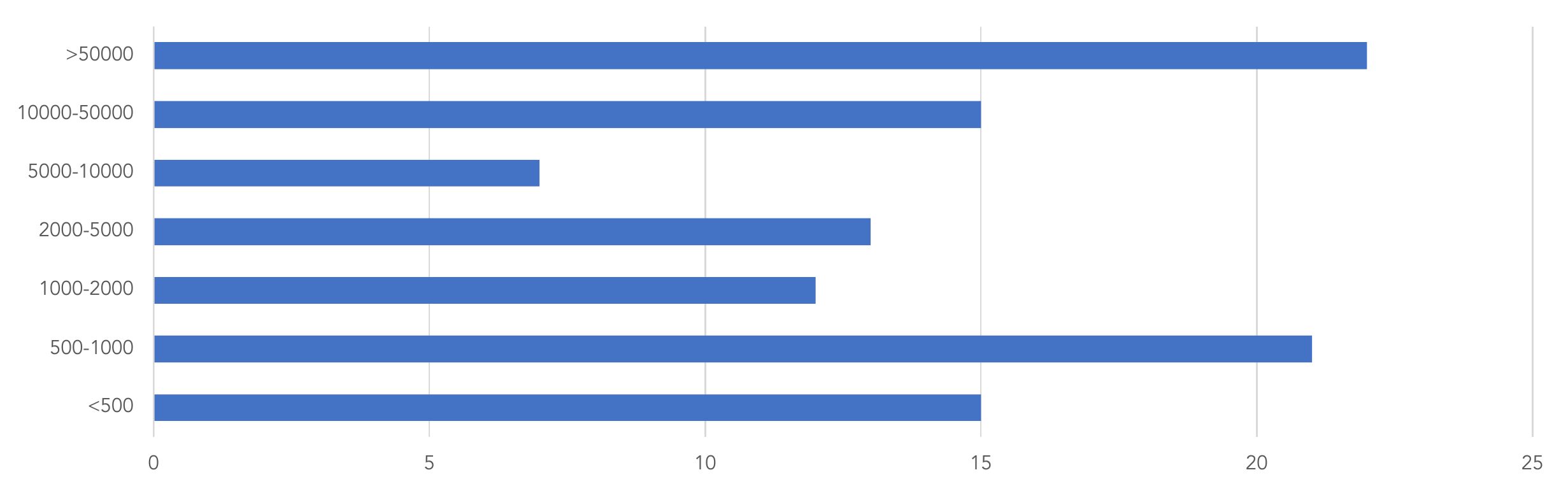}
    \caption{The result of Q13 - The number of documents needed to build the knowledge base.}
    \label{fig:document-numbers}
\end{figure}

\subsubsection{Geoscience knowledge base construction is a complex multi-stage, multimodal data annotation task}\label{Task}
Based on the results of the questionnaire shown in Figure \ref{fig:document-numbers}, when the research team work on database construction, they generally believe that a very large number of articles are required to collect enough data to drive research.
However, at the same time, they don't have a large team size and often face the situation where a team of 10-15 people need to process thousands of documents.
The result also shows that the databases they need to build generally have 100-150 attribute fields, and we learned more about the composition of the database in our in-depth interviews.
Some interviewees also mentioned this situation:

\textit{``The total number of articles in my research field should be around 200,000, and I need to filter out articles that contain data. I think there should be tens of thousands.''}-PA01

According to the interviewees, most scientific knowledge bases in the field of geology often store the measurement data of rock samples, and the attribute type of the field is often the multiple analysis and measurement of the sample, including the chemical element content analysis of the sample, the specific information of the sample collection point, the lithology information of the sample, and the meta information of the article.

\textit{``Different researchers have different forms of expression in the data they focus on because of their different perspectives on research questions.''}-PA05

For example, the geographic location information of the sample collection point may exist on the map figures, but the specific location description may exist in the text or the figure's caption. The same problem arises when it comes to lithology and age-related data. 
Authors will have differences in the way of writing when describing the address object, and often the articles published by different teams have differences in the way of related descriptions.
This situation leads them to process and extract image information, table information and text information respectively for a document.

\begin{figure}
    \centering
    \includegraphics[width=\linewidth]{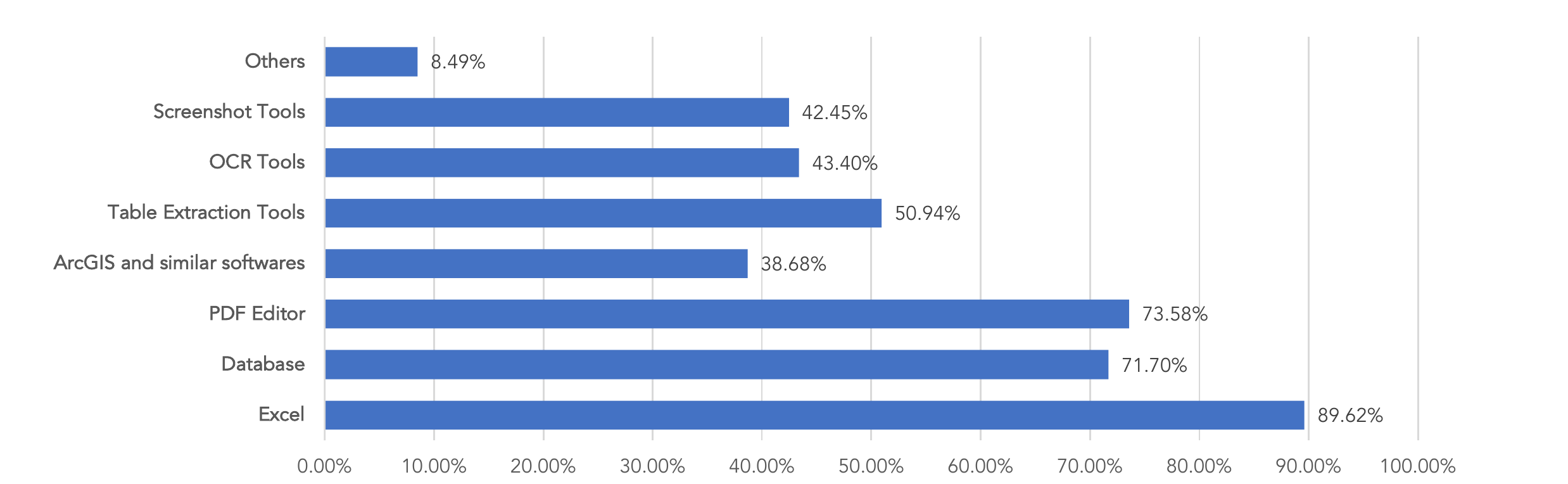}
    \caption{The results of Q7-Tools used in data extraction (more than one answer is possible) shows that the researchers are using lots of tools to process the PDF. We also found that except for the tool for storing structured data, the most used tool is PDF Editor, which shows that the way they currently process documents is still relatively simple and still relies on its own operations without resorting to batch processing tools.}
    \label{fig:tools_result_questionnaire}
\end{figure}

The multimodal form of the data forces the data extraction process to be split into multiple stages.
The interviewed participants mentioned that they did not have one tool to assist them in this multimodal information extraction process. 
This situation is also reflected in the results of the questionnaire.
As shown in Figure \ref{fig:tools_result_questionnaire}, researchers often use a combination of a series of tools.

In this paper, we focus on the data extraction and linking process of their research.
Based on the results of interviews and questionnaires, we described the scientific knowledge base construction process and defined the tasks in the process:
\begin{itemize}
  \item \textbf{Step 1-Meta Data Record:} Extract metadata of articles for subsequent traceability of data sources,
    \item \textbf{Step 2-Data Existence Confirm:} Quickly browse the article to find the table that describes the key entities (often samples or geological objects depend on the fields),
    \item \textbf{Step 3-Table Extraction:} Get the data of the key entities from the table and fill in the Excel file prepared in the advance cell by cell,
    \item \textbf{Step 4-Data Completion:} 
        \SubItem{\textbf{From text:}} Search the full text with keywords to locate the attributes of the key entities (for example, the lithology of geological objects), fill in the data in the Excel file after finding it, and repeat until all the data is found,
        \SubItem{\textbf{From figure:}} Restore the corresponding information from the figures (for example, obtaining the latitude and longitude of a marked point from the map),
    \item \textbf{Step 5-Proofreading:} Check and proofread to ensure the data is accurate,
    \item \textbf{Step 6-Reference Linking:} Record the article reference information and assign a Reference ID, and link the data to the corresponding Reference ID,
    \item \textbf{Step 7-From Excel to Database:} Integrate the data extracted from each article (usually stored in a bunch of Excel files) into the database.
\end{itemize}

\subsubsection{Attempts at crowdworker-based solution are not successsful due to the lack of highly professional domain knowledge}

Our interview participants also mentioned that their teams try to hire some full-time data extraction engineers with non-professional backgrounds to do data extraction due to the heavy burden of data extraction.
We interviewed the full-time data entry assistants (PA11 and PA12) and found that because they do not have a professional background in geoscience, it takes a long time to conduct relevant knowledge training before starting the data extraction work.

\textit{``We trained for nearly a month to learn what each field in the knowledge base means and where this data will appear in the article. In addition, we also need to learn relevant domain knowledge to understand the article.''} -PA12

The more serious problem is that when they move to another research topic, the training needs to be redone due to changes in the types and attributes of the data involved, which proves that full-time data extraction workers cannot be used for data extraction with domain knowledge.

\textit{``But when we did data extraction in the other research topic, we had to relearn from scratch because the knowledge we needed was completely different.''} -PA11

On the one hand, this situation reflects that full-time data extraction workers without domain knowledge do not reduce the difficulty of data extraction, and additional training is required to help workers become familiar with the field, which may bring additional costs.
On the other hand, it also reflects the same problem that might happen when obtaining the training data required for the human-in-the-loop model training process.
The requirements of highly professional domain knowledge in the scientific knowledge base construction task lead to the result that only domain experts can provide high-quality labeled data: domain experts will be still trapped in the heavy data annotation task.

\subsubsection{The exploration of fully automatic solutions with human-in-the-loop model training also faces various difficulties}

We learned from the formative study that many teams had tried an end-to-end approach to building knowledge bases. 
The main reasons for the failure are that domain experts lack technical knowledge and have a high requirement for data accuracy.

From the survey research and in-depth interview, we conclude that the typical profile of our users is an average computer user with no / low programming skills.
Only PA07 is a proficient programmer and can use Python scripts to process the data, but he still mentioned that

\textit{``Being unfamiliar with machine learning makes me unable to deal with the data in pictures and charts in the papers efficiently. I am not sure whether a toolkit can help me extract all the pictures from PDFs''.}

\begin{figure}
    \centering
    \includegraphics[width=0.95\linewidth]{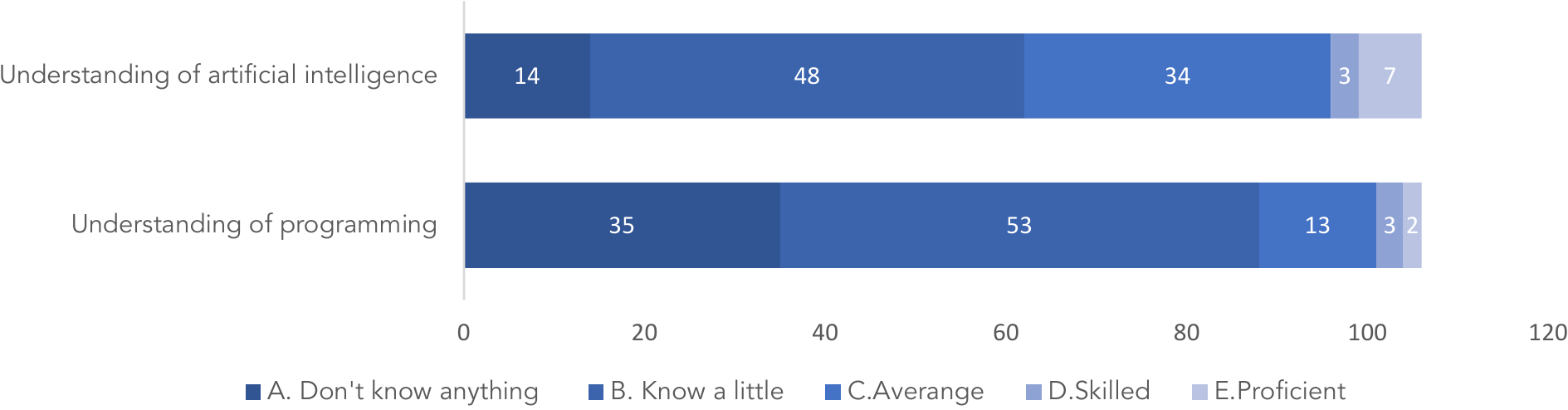}
    \caption{The results of Q18-Understanding of programming and Q20-Understanding of artificial intelligence, which show that the vast majority of researchers in our sampled group currently have a low level of practical understanding of programming and artificial intelligence.}
    \label{fig:understanding_of_ML}
\end{figure}

We can find that most participants of the survey also lack an understanding of programming/artificial intelligence (Figure \ref{fig:understanding_of_ML}).
Furthermore, some interviewees also report that they always encounter difficulties in data extraction and processing due to their lack of programming skills.

The questionnaire results also reflect that some teams want or are already trying to cooperate with computer science experts to use some end-to-end tools to process the data obtained, but still have dilemmas:

\textit{``The computer science students who help with the data processing don't quite understand what we need to do with the data.''}-PA06

It reflects that big-data-driven researches require interdisciplinary cooperation in the data collection process. However, there are still some communication problems between domain experts and data experts on data processing issues, which is also consistent with the results of previous studies \cite{10.1145/3290605.3300356}.

\subsubsection{AI can not meet the high precision requirement for the data extraction task in the scientific knowledge base construction}
Another major challenge is that all interviewees indicated that they need the  extracted data to be very accurate, and each piece of data needs to have a clear source and be double-checked before it can be stored in a scientific knowledge base.
They agree that the current end-to-end model does not meet their requirements.

\textit{``I think the data processed end-to-end still requires much manpower for proofreading.''} -PA01

Some participants mentioned that they tried the human-in-the-loop method and think that the time and labor costs they pay are not much different from manual extraction.

\textit{``I asked machine learning experts how to train a model for extraction and learned that it takes a lot of labeled data to achieve the accuracy I envisioned. I think the cost of labeling data may be similar to the cost of direct manual extraction.''} -PA10

It shows that some of the current opinions of the domain experts on the human-in-the-loop approach are not positive, and they are not entirely convinced that this method can reduce their burden on the premise of ensuring accuracy.

We found that domain experts have a certain degree of misunderstanding about the tasks that AI can accomplish due to the lack of understanding of AI technology, which also caused their expectations to differ from the actual human-in-the-loop machine learning implementation route.
We can think of this as the difference between a domain expert's mental model of AI technology and an actual conceptual model.
The discrepancy between the great effort annotations and the resulting accuracy was considered unacceptable by domain experts, which was also reflected in the fact that many respondents repeatedly mentioned their need for 100\% accuracy.

\subsubsection{Summary of Key Insights}
The key insights we obtained show that building a scientific knowledge base requires the participation of experts to ensure extremely high data accuracy. 
However, due to domain experts' lack of understanding of computer-related technologies, a graphical interface is needed to assist in the cooperation between experts and AI. 
Therefore, we believe that the form of AI-in-the-loop is the most suitable for this scenario.
The AI-in-the-loop approach allows experts to effectively control the accuracy of the extracted data while using AI assistance to reduce workload.
After considering the AI-in-the-loop approach, we propose the initial design of \KBS~framework as a design probe to solicit users' feedback.

\section{Participatory Design}

We conducted participatory design (PD) sessions \cite{kuhn1993participatory} to further iterate and uncover design opportunities for \KBS.

We recruited 14 researchers (PB01-PB14) from 9 different research groups (GB1-GB9) from geoscience departments of different universities. The details are provided in Table \ref{tab:participatory_participants}.

\begin{table}
    \centering
    \begin{tabular}{c|c|c|l}
    \toprule
    Group ID &  Participant ID  & Gender & Identity \\
    \midrule
      GB1   &  PB01 & Male & PhD Student\\
      GB1   &  PB02 & Female & PhD Student\\
      GB1   &  PB03 & Male & PhD Student\\
      GB2   &  PB04 & Female & PhD Student\\
      GB2   &  PB05 & Male & Associate Professor\\
      GB2   &  PB06 & Male & PhD Student\\
      GB3   &  PB07 & Male & Postdoctoral Researcher\\
      GB4   &  PB08 & Female & PhD Student\\
      GB5   &  PB09 & Female & PhD Student\\
      GB6   &  PB10 & Male & PhD Student\\
      GB6   &  PB11 & Female & PhD Student\\
      GB7   &  PB12 & Male & PhD Student\\
      GB8   &  PB13 & Female & PhD Student\\
      GB9   &  PB14 & Female & Associate Researcher\\
    \bottomrule
    \end{tabular}
    \caption{Demographics of Participatory Design participants.}
    \label{tab:participatory_participants}
\end{table}

\subsection{Procedure}
Each session lasted around 45 minutes and was conducted remotely via Tencent Meeting due to the COVID-19 pandemic. 
Due to the differences in the research topics of the participants, all PD sessions are grouped by the research group of the participants, and participants from the same research group have the PD session together.
All sessions are recorded on video.
The participants were asked to share their feelings of the prototype and give some possible design opportunities and new features to improve the current design. 
Before each PD session, we conducted semi-structured interviews with each group of participants to understand their current specific research topic and the structure of the scientific knowledge base that needs to be constructed.
Then in the PD session, we provided an online-accessible prototype, along with the functional ideas for each part.
Participants accessed \KBS~prototype using the browser on their computers and were asked to share the screen with experimenter.

\begin{figure*}
    \centering
    \includegraphics[width=\linewidth]{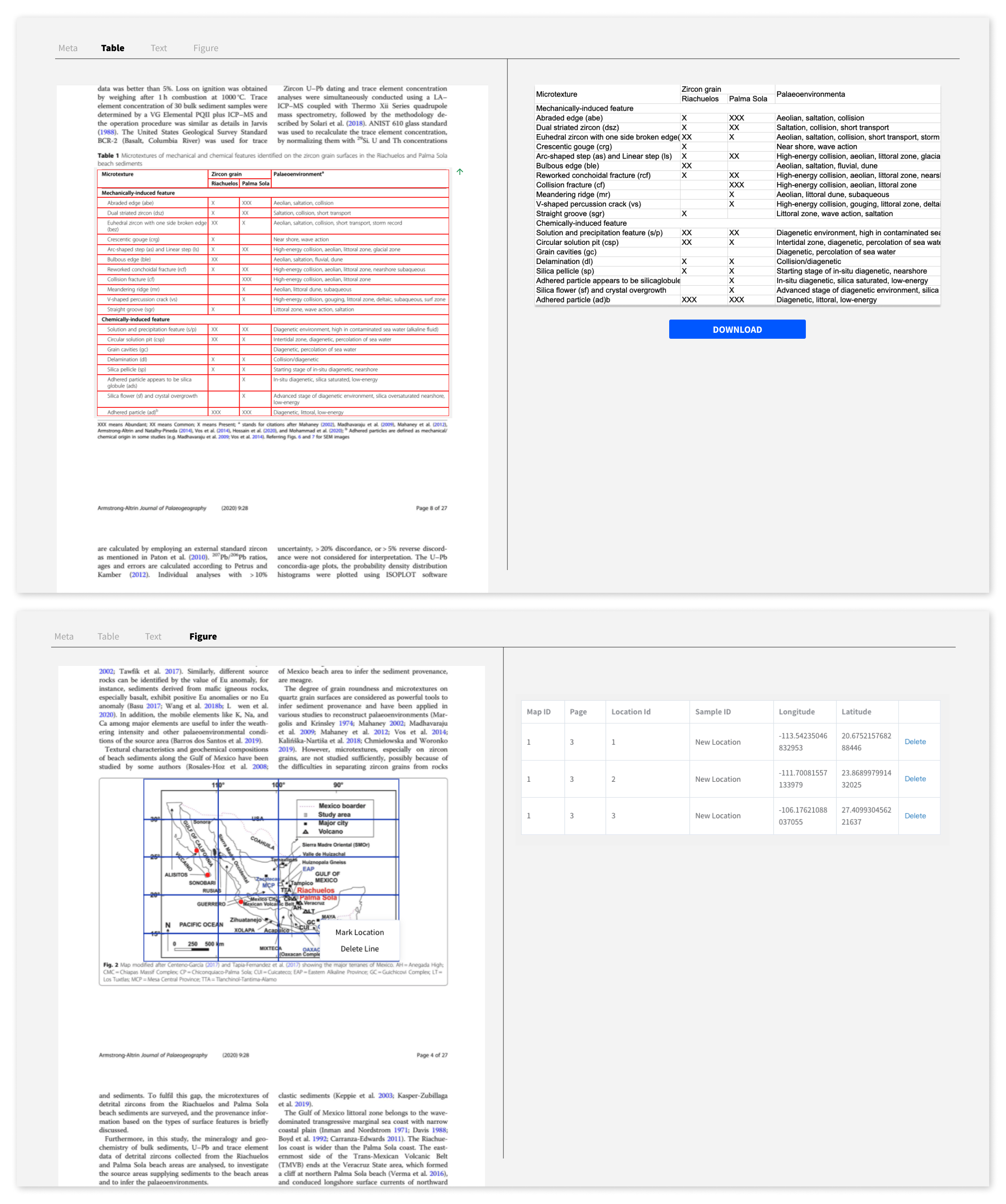}
    \caption{The prototype used in the PD session. }
    \label{fig:prototype}
\end{figure*}

\subsection{Prototype}
The design of the prototype (see Figure \ref{fig:prototype}) used in the PD session was derived from the conclusions of the formative research.
We used Vue.js and FastAPI to implement a simple interactive prototype and deployed it for implement of the remote PD sessions.
This prototype contains our initial implementation of the AI model, so all participants can try the prototype for a simple attempt at data extraction.
The prototype system can show the results of meta information extraction (Step 1 in Section \ref{Task}) and provide a table extraction (Step 3 in Section \ref{Task}) function and a figure information extraction (Step 4 in Section \ref{Task}) that can interact with the model.

During the interaction, we asked participants to think about the connection between their current workflow and the one shown in the prototype. We focus on whether our process design and functional design can meet the current requirements of researchers to build scientific knowledge bases, including requirements for efficiency and data accuracy.
We hope that as participants interact with the prototype, participants will be able to provide insights into the design of key features, including how they feel, and whether they think the similarities and differences with current workflows will negatively impact their workflows.
Through this process, each participant can clearly understand which functions are effective in improving their own work efficiency. Going a step further, we asked them to think about other functional design ideas, including (1) bridging the gap between the process in the existing prototype and the current process; and (2) addressing the functional flaws they found that still existed.

\subsection{Findings}

\subsubsection{More precise step-by-step interaction with the model helps control the accuracy of the data}\label{data-accuracy}
All the participants consider errors unacceptable, as they will directly impact the reliability of follow-up research results.
Taking the table as an example, the tables in PDF are usually not reproduced very well by ordinary PDF editors, which makes data collection very difficult.
Several participants told us they would spend a lot of time extracting tables because they had to ensure the numbers in each cell were correct. 
PB06 mentioned that since the manual process is directly copied or entered the data to the excel files, it is easy to make mistakes.

They agreed with the table extraction process we showed, which they believed would help them extract data faster and avoid other errors introduced manually.
Moreover, they believe that the structure and content of the table can be further split to mitigate errors better.
The same conclusion applies to the process of extracting latitude and longitude from figures, during which the participants hope to correct some minor errors to make the results more accurate.
At the same time, they also agree that with this method they can understand what tasks the model can accomplish  more intuitively.

\subsubsection{Integrating data from the results of different steps is needed}\label{intergaration}
During the discussion, the participants also proposed ideas about the final form of the data. 
Due to the multi-modality of the original data, after the data extraction part is implemented, the data is still scattered and divided into different modules for storage.
The final database must link all the scattered data to realize query and analysis.
If only data from different parts are exported separately, researchers need to integrate the data manually.
PB10 believes that the current extraction functions can achieve the extraction effect, but there is still a certain distance from the final database.
They felt that if \KBS~has the ability to incorporate automatic data integration from the single document level to the project level, it would significantly reduce the burden of their manual integration.
PB01 from GB1 said, ``The integration of data from a single table into a comprehensive table might allow us to enter all the data into the database faster''.

\subsubsection{Research teams need customized knowledge base building capabilities}\label{granularity}
In the PD session, we corroborated findings from formative research that there is some variation in the data granularity that different teams need to extract.
There are significant differences in the types and amounts of data that different research fields focus on.
For example, PB07 mentioned that the data he is concerned about only exists in the text description, and he only needs to process the text part and does not need to extract the table.
Most data are concentrated in the table in some research fields, and some supplementary attributes will be found in other parts.
It leads to differences in the data extraction tasks that need to be completed in the actual extraction process, and some functions may not be used at all during the construction of some scientific knowledge bases.

While testing the prototype, several participants mentioned that their team worked on more than one scientific knowledge base. They felt they desperately needed a file management system to help them organize their different stages of database builds.
PB01 said they have 18 people on the team, but the situations are mainly different when they are in a minor team.
The strategy adopted by some teams is to work together to build a database. In contrast, some teams have different members responsible for different databases construction, requiring various configuration solutions to adapt to different scientific knowledge base construction forms.
Moreover, multi-project teams will have different database data table structures, which affects the data integration process and puts forward requirements for customized configuration of data integration functions.

\subsection{Design Strategies}
We conclude the 4 following design strategies from the PD sessions, which guide us in the system design:
\begin{itemize}
    \item \textbf{DS1: }Guarantee the decision-making power in the data extraction process and reduce errors in the interaction process with the model to ensure the accuracy of the final extracted scientific data.(Section \ref{data-accuracy}).
    \item \textbf{DS2: }Use automatic data integration at the document and project levels to replace manual integration to reduce the burden on experts.(Section \ref{intergaration})
    \item \textbf{DS3: }Establish document and project management processes for multiple databases construction. (Section \ref{granularity})
    \item \textbf{DS4: }Diversified project configurations to accommodate data extraction tasks of different granularities, including the format of processing files and the configuration of extraction functions.(Section \ref{granularity})
\end{itemize}

\section{\KBS}
Following these four design strategies, we designed and implemented \KBS, an AI-in-the-loop system for building a scientific knowledge base from the literature that helps researchers intuitively process literature and extract data.
\KBS~aims to introduce AI into the current human workflow to improve human efficiency and data accuracy with a multi-step multi-modal human-AI collaboration pipeline .
The system implementation details are shown in Appendix \ref{data_extraction_model}.

\subsection{System Overview}
\begin{figure}
    \centering
    \includegraphics[width=0.95\linewidth]{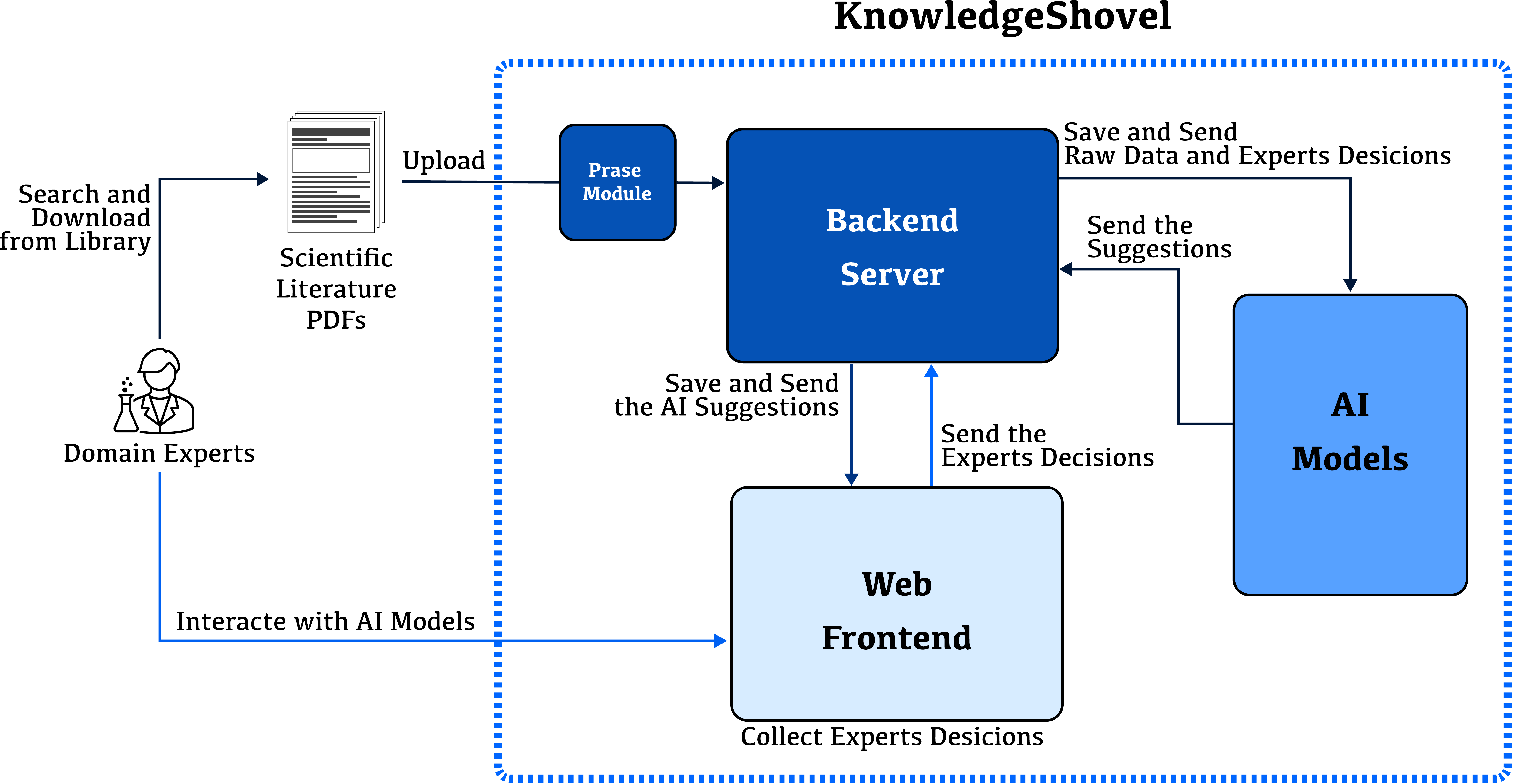}
    \caption{\KBS~System Overview}
    \label{fig:system_overview}
\end{figure}
As shown in Figure \ref{fig:system_overview}, \KBS~consists of: 
 (1) an interactive front-end graphical user interface including document management and user system (Appendix \ref{documentmangement}), single-document-level data extraction (Section \ref{single}) and project-level data integration (Section \ref{project-level}); 
(2) a back-end parse module to pre-process the PDF format files and extract metadata;
(3) a back-end server for processing user-model interaction data and supporting file management and user systems; 
and (4) several back-end neural network models supporting data extraction and integration functions.

\subsection{Single Document Process}\label{single}
\begin{figure}
    \centering
    \includegraphics[width=\linewidth]{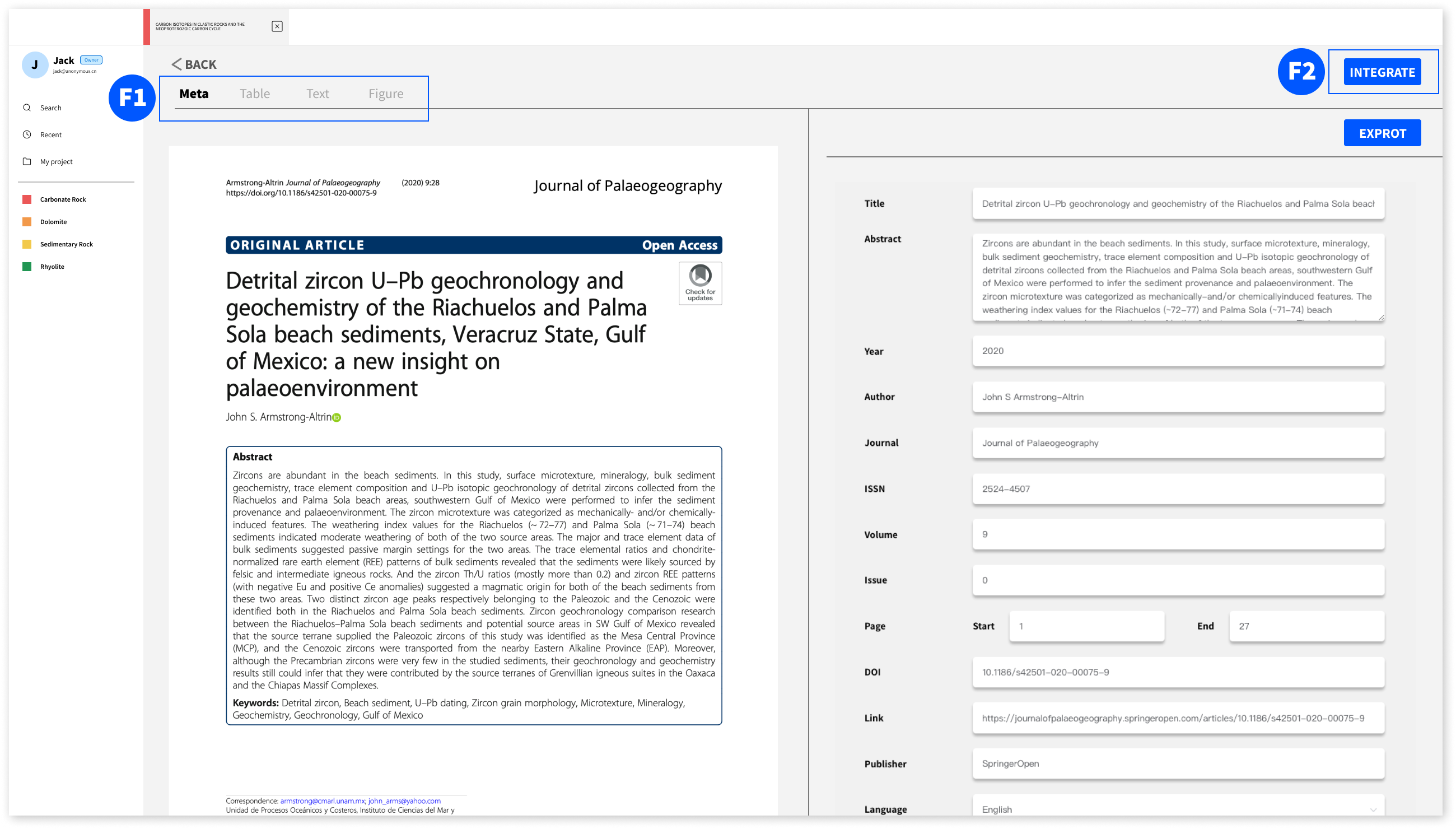}
    \caption{Metadata Extraction Function in \KBS. The figure shows the metadata of the article \textit{``Detrital zircon U–Pb geochronology and geochemistry of the Riachuelos and Palma Sola beach sediments, Veracruz State, Gulf of Mexico: a new insight on palaeoenvironment}'' \cite{armstrong2020detrital}.}
    \label{fig:meta}
\end{figure}

When users open a file from Project File List to start their work, they will enter the data extraction interface (Figure \ref{fig:meta}). 
In the data extraction interface, users can switch the different tabs (e.g., Meta, Text, Table, and Map) in the area F1. Each extraction workflow follows a multi-step human-AI collaboration pipeline that helps users interact with the model.

According to the results of our formative study, researchers need to obtain the meta-information of the article to trace the data.
For the task of extracting the articles' meta information, we develop a module to automatically extract the title, author(s), journal/conference, and other meta information from the PDF file in the \textbf{Meta} tab. 
As shown in Figure \ref{fig:meta}, users can edit and save the meta information that can be joined to the output dataset.
We used multiple parsing toolkits to extraction the metadata, the implementation details are in Appendix \ref{extraction-implementation}.
This function can help users save the data source in the database to trace back and cite these articles in subsequent research.

\begin{figure}
    \centering
    \includegraphics[width=\linewidth]{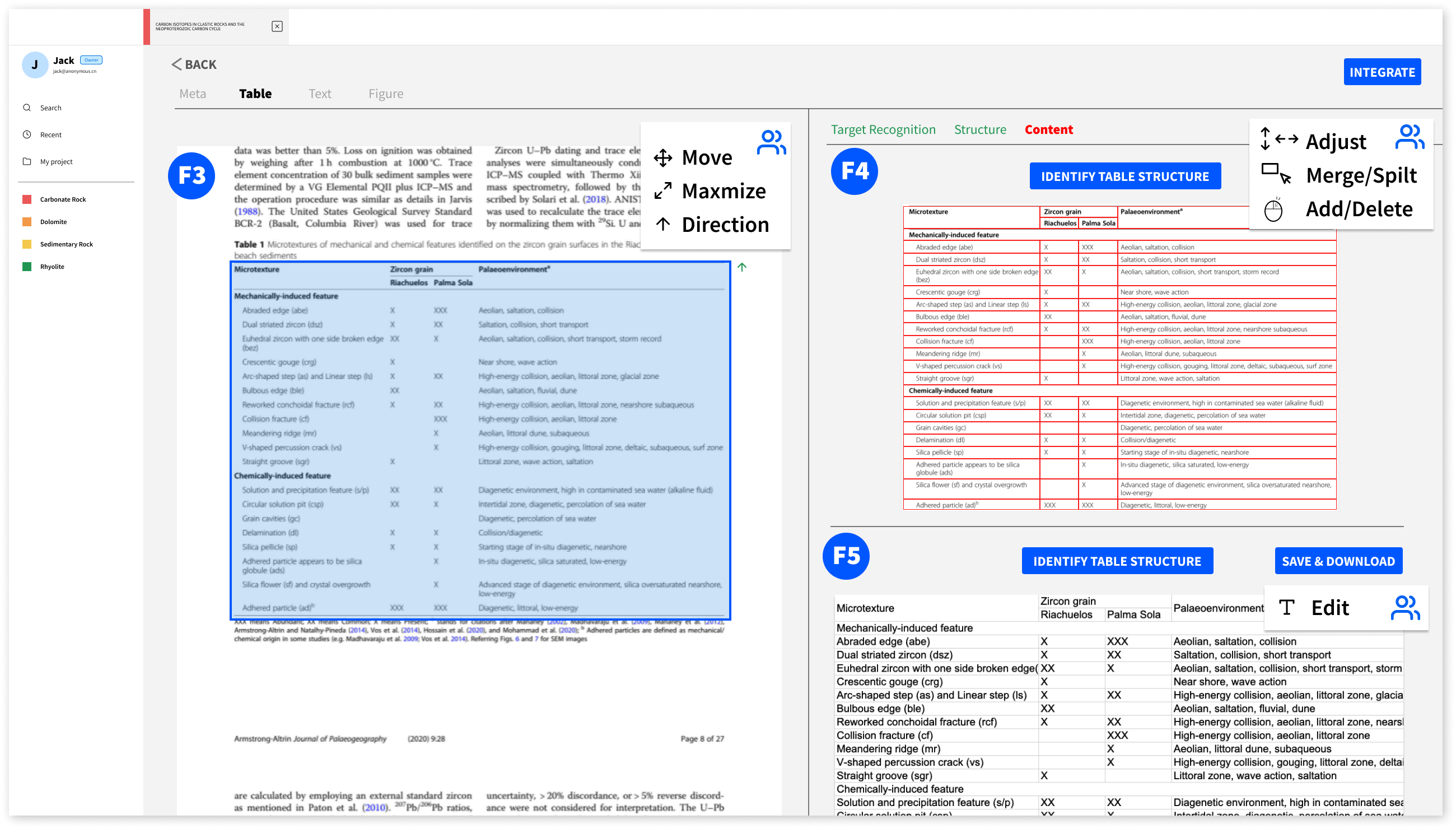}
    \caption{Table Extraction Function in \KBS.}
    \label{fig:table}
\end{figure}

From the results of PD and what we have learned from the formative study, most of the data exist in the form of tables in the article, and table extraction is the core and most arduous task of building a scientific knowledge base. 
We develop the table extraction function in the \textbf{Table} tab (Figure \ref{fig:table}).  
In this part, the human-AI collaboration pipeline has the following 6 steps: 
(1) AI model pre-located the position of each table in the PDF;
(2) User chooses a table and confirm the table area;
(3) AI model recognizes the table's structure;
(4) User gets the table's structure result from AI model and confirm;
(5) AI model recognizes the table's content;
(6) User edits and confirms the content.
In each step, the back-end models help people to easily get the result and collect the users' adjustments, which can be a dataset for model training and fine-tuning (see Figure \ref{fig:table_extraction_interaction}). 
In the formative study, both PA02 and PA05 mentioned that the accuracy of the data in the complete end-to-end extraction and processing of the table could not meet the requirements of scientific knowledge bases, but manual processing of the table is very tedious work.
The human-AI collaboration pipeline in \KBS~can obtains the data quickly with AI assistance, and whether the data is correct is still up to the human decision, which ensures the accuracy of the data to the greatest extent and reduces a large number of steps compared to complete manual cell-by-cell data copying.
From the user's perceptive, the first step is adjusting where the table is in the F3 area or drawing a new area as a table, then starting to recognize the structure.
The next step is to adjust the structure that the system advised (F4).
The system provides `add and delete column/row' and `merge or split cell function.'
After structure recognition, users can start the content recognition and edit the content in each cell (F5).

\begin{figure}
    \centering
    \includegraphics[width=\linewidth]{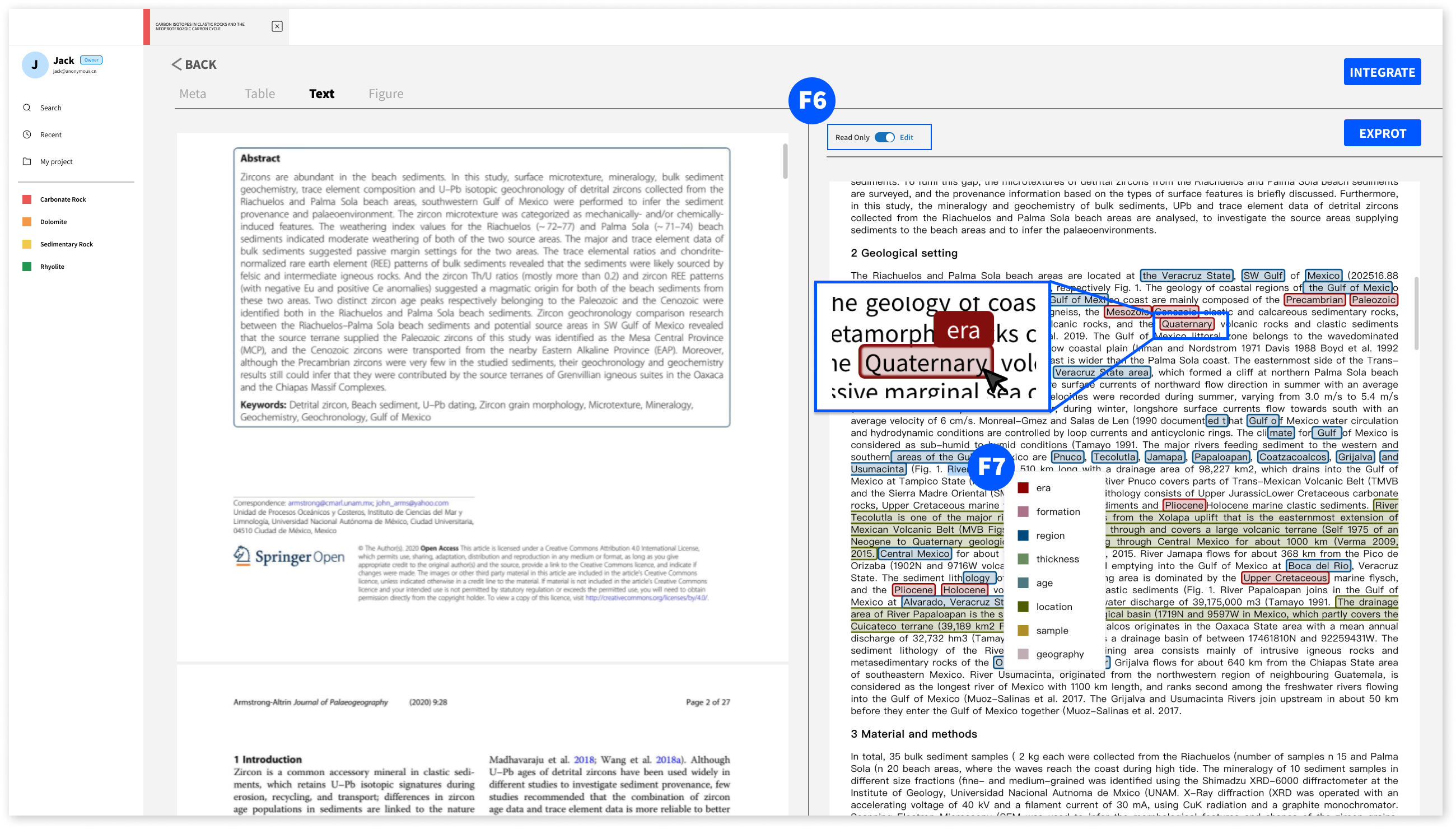}
    \caption{Text Extraction Function in \KBS.}
    \label{fig:text}
\end{figure}

\begin{figure}[!t]
\centering
\subfloat[The settings of text extraction.]{\includegraphics[width=0.48\linewidth]{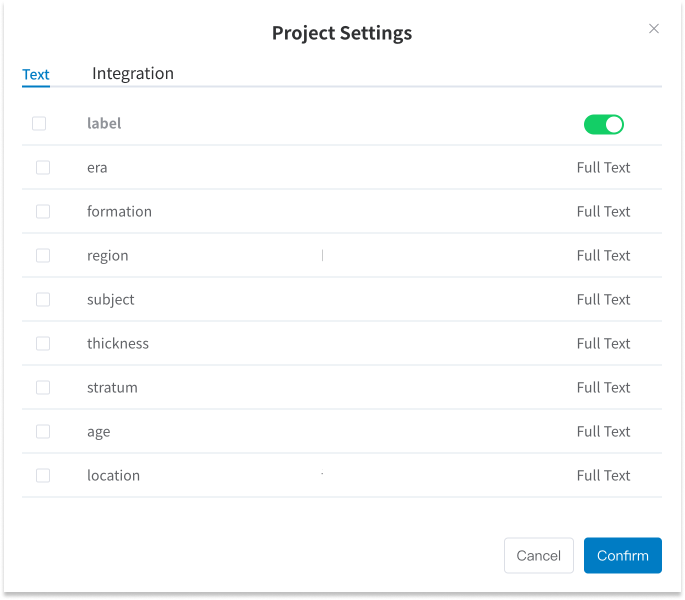}
\label{fig:table_fusion_setting}}
\hfil
\subfloat[The settings of data integration.]{\includegraphics[width=0.48\linewidth]{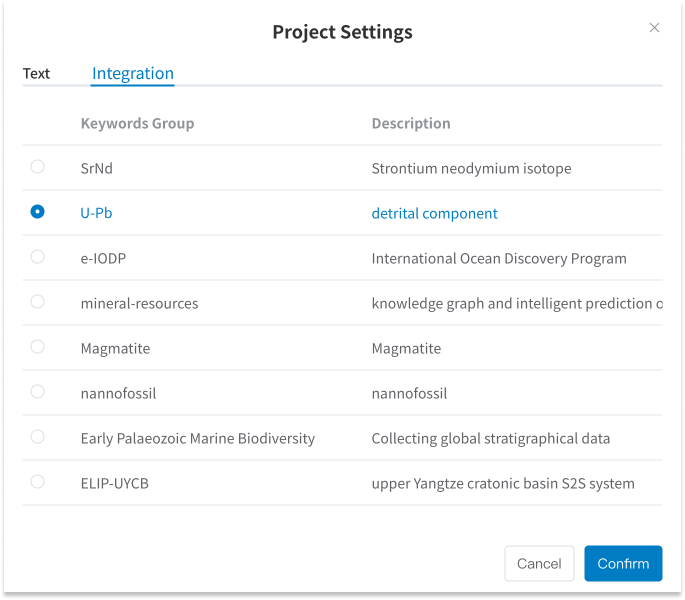}
\label{fig:text_setting}}
\caption{The settings for extraction and integration.}
\label{fig:settings}
\end{figure}

In the \textbf{Text} tab, the weak-supervision learning models and rules are used to finish a named entity recognition task.
The result can help highlight the focused keywords in texts and the samples' attributes to help them add these words into the datasets.
Users can choose to show or hide some labels set at the project level, as shown in Figure \ref{fig:text_setting}.
All pre-extracted entities are highlighted so that users can quickly locate information.
Users can also annotate a keyword via mouse selection when they switch to the edit mode (F6) and select a label (F7), as shown in Figure \ref{fig:text}. 
User can use the right-click to delete the wrongly extracted words.
The labeled word can be added to the output database when linked to a designated scientific entity (which should be set in the project settings).

\begin{figure}
    \centering
    \includegraphics[width=\linewidth]{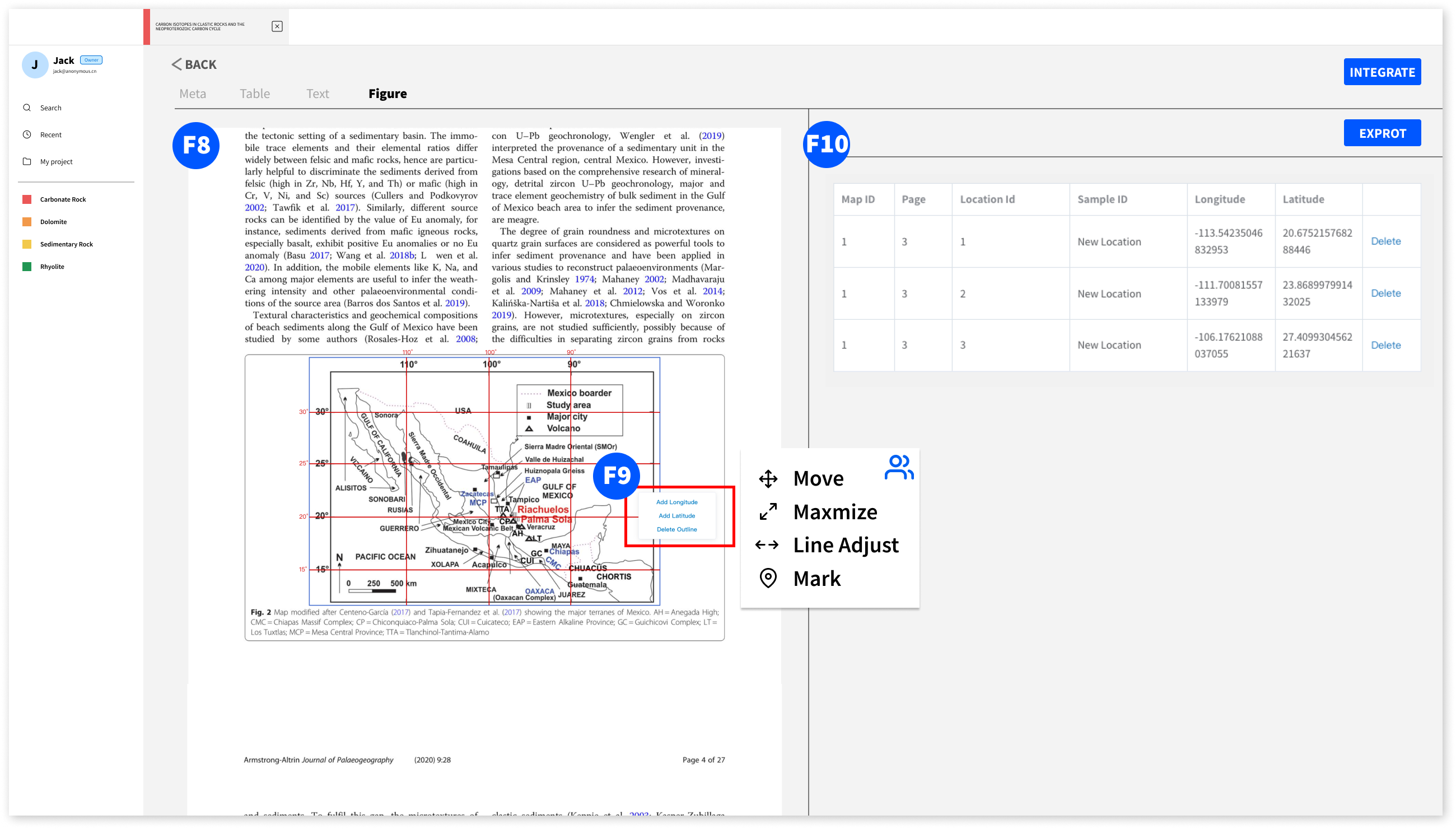}
    \caption{Figure Information Extraction Function in \KBS.}
    \label{fig:map}
\end{figure}

The \textbf{Figure} tab is for collecting the location of a sample.
We provide a module that can recognize maps and calculate the latitude and longitude of each point on the map (Figure \ref{fig:map}). 
Users can draw an area (F8) that contains the map and adjust the recognization result by adding new line and editing the value (F9).
Then users can mark a point by right click. The latitude and longitude will automatically be saved in the table (F10) as shown in Figure \ref{fig:map} and can be joined to the output database.
Similar to the table extraction, the human-AI collaboration pipeline has the following 5 steps:
(1) AI model pre-located the position of each map in the PDF;
(2) User chooses a map and confirm the map area;
(3) AI model recognizes the map's structure;
(4) User gets the map's structure result from AI model and confirms;
(5) Users mark the point for final data.

From the results of participatory design session, participants would like to obtain more detailed interactive operations.
Therefore, \KBS~allow users to modify the identified latitude and longitude values and the position of the latitude and longitude lines in the processing of the map, which can also help to collect more accurate results.

\begin{figure}
    \centering
    \includegraphics[width=\linewidth]{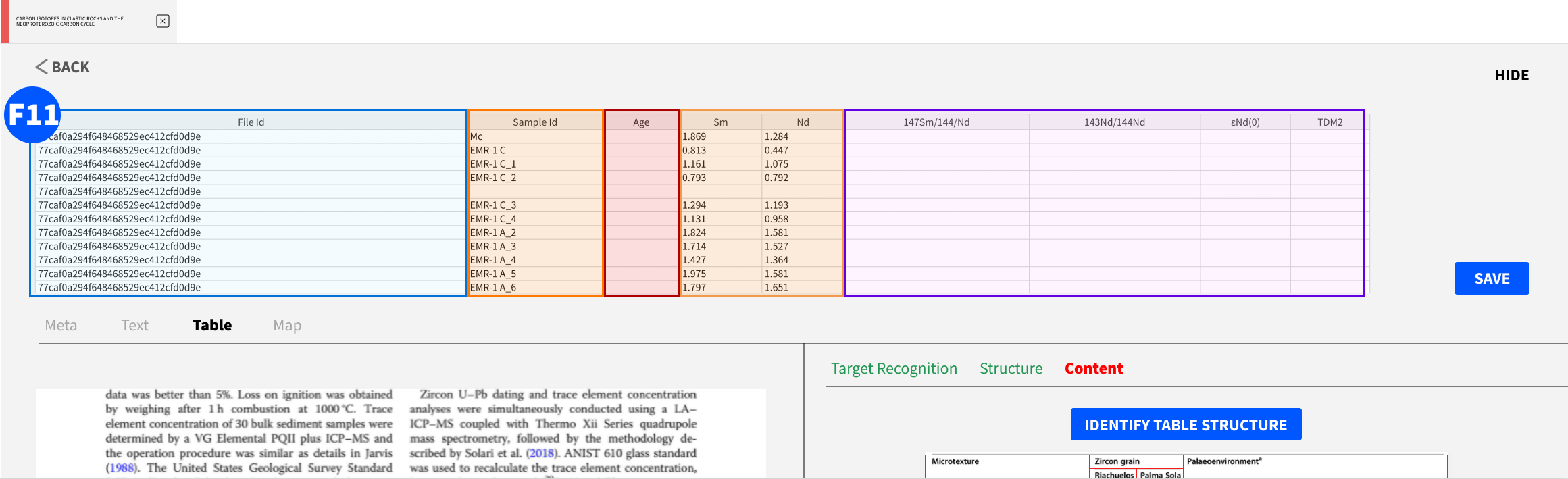}
    \caption{The integration function for a single document. \textbf{\textcolor{blue}{Blue}}: Metadata ID. \textbf{\textcolor{red}{Red}}: Information from text. \textbf{\textcolor{orange}{Orange}}: Information from a table, including the key entity ``Sample ID'' and its attributes.  \textbf{\textcolor[RGB]{96,0,227}{Purple}}: Information from the other table including key entity's attributes.}
    \label{fig:integration}
\end{figure}

After the data is extracted step by step, it needs to be integrated into a table to establish a database. 
The user needs to set the header of the table used for integrating on the project page (Figure \ref{fig:table_fusion_setting}, we detailedly introduce in Section \ref{project-level}).
Then in the data extraction interface (Figure \ref{fig:meta}), when users click the Integrate button (F2), the back-end model will process the data in each part, including metadata, tables, and location in maps and texts. 
The result is shown in the F11 area in Figure \ref{fig:integration}.
Users can save and download this document-level result, which also is joined to the project-level result by the user selected.

\subsection{Project Level Process}\label{project-level}
Insights from PD sessions indicate that researchers often need to integrate data from multiple documents. 
Based on this, we designed the project-level data integration process. 
This process is based on the data table header settings that the user configures for the project.
We provide the function of uploading and automatic parsing to simplify the input process. 
Users can directly upload an existing excel file to create a header configuration quickly.
Based on feedback from participants in the PD session, we allow users to remove or add some header items after uploading to accommodate their possible changes in the data extraction process.
After this configuration is complete, this configuration will be applied to all single documents in the project.

After this configuration item takes effect, all user-saved extracted data in tables, texts, and images will be automatically integrated into a single file, becoming the basis for project-level data integration.
Users can integrate the summary table of each file into a project-level summary table at the Files List interface, and the result can be automatically integrated into an excel file for download.
The final exported file will contain the reference information of each file and all user-saved data.

\subsection{Document Management and User System}

We designed the projects list interface to present the pertinent details of each project in order to implement project management.
Each project has a file list that details who uploaded each file, who edited it most recently, when each was uploaded and edited, and whether the file has a principal.
In order to prevent several users from working on the same file at the same time, we created a file locking mechanism that takes into account the specificity of system operations and the interaction process with the back-end models.
We further implement a principal mechanism based on file lock.
Users can become the "principal" in control of a file by clicking the "Take Charge" button in the list.
The project settings include the text labels, the export dataset headers, and the project description.
Considering that the dataset may contain several headers, we provide batch editing for convenience.

\section{User Study}

From February 2022, \KBS~has deployed, and more than 40 geoscience research teams from geoscience departments of more than 10 universities are invited to use \KBS~for data extraction and scientific knowledge base construction. 
After 6 months, we conduct a user study to evaluate \KBS.

The user study focuses on the following research questions:
\begin{itemize}
    \item \textbf{RQ1: Can researchers successfully obtain the data that can build the knowledge base?}
    \item \textbf{RQ2: Can \KBS~improve the efficiency and data quality in the scientific knowledge base construction?}
\end{itemize}

\subsection{Participants}
We recruited 7 student users from our users to launch the user study. 
The demographic characteristics of the participants are reported in Table \ref{tab:evaluation_participants}.
Each participant received ¥400 for their participation.

\begin{table}
    \centering
    \begin{tabular}{c|c|c|l}
    \toprule
    Participant ID &  Gender & Age & Identity \\
    \midrule
      PC01   & Male   & 20 & Undergraduate Student\\
      PC02   & Female & 22 & Graduate Student (Master)\\
      PC03   & Male   & 22 & Graduate Student (Master)\\
      PC04   & Male   & 21 & Undergraduate Student\\
      PC05   & Male   & 20 & Undergraduate Student\\
      PC06   & Female & 24 & Graduate Student (Master)\\
      PC07   & Female & 26 & Graduate Student (PhD)\\
    \bottomrule
    \end{tabular}
    \caption{Demographics of Evaluation participants.}
    \label{tab:evaluation_participants}
\end{table}

\subsection{Procedure}
\begin{figure}
    \centering
    \includegraphics[width=0.7\linewidth]{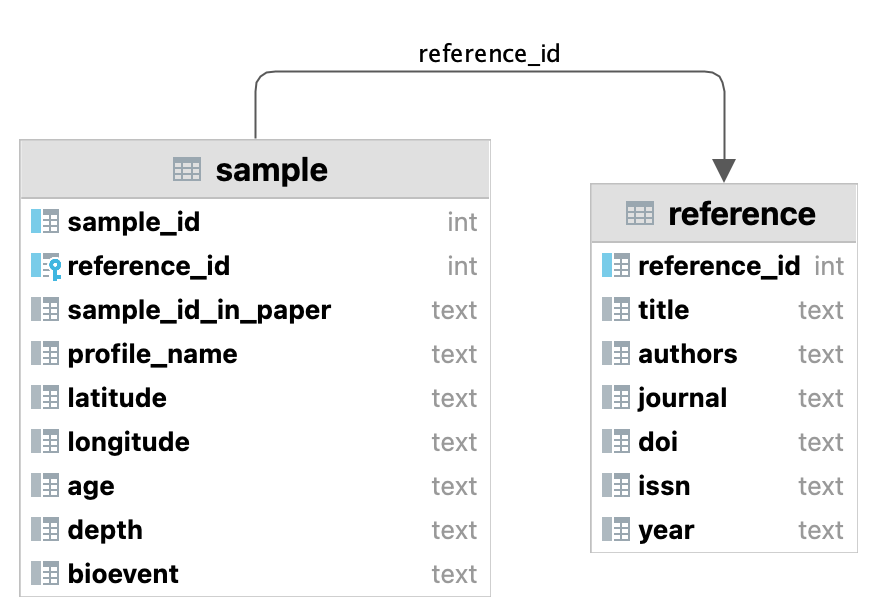}
    \caption{The UML of Age Model database we test in user study session. The Age Model database is used to build the age model for ocean drilling project including DSDP ODP and Two IODPs \cite{iodp}.}
    \label{fig:database-schema}
\end{figure}

Each user study session lasted around 2 hours.
We visited the research team in person and conducted the user study session in their office.
Considering the participants are all the current users of \KBS, the experimenters only have 5 minutes to briefly introduce the \KBS~and we also give a 5-minute introduction of our study and collect the participant's demographic characteristics.
In order to evaluate the system coverage of extraction tasks, the participants are asked to extract the data of 3 representative articles about their research which is for building the Age Model database (the used dataset's schema is shown in Figure \ref{fig:database-schema}). 
The detailed meta information of the article is shown in Table \ref{tab:articles}.
The selection of these three articles was recommended by the PI of the participant's research team and is in line with the research topic of the team. 
In order to control the research time, the three articles each contains under 30 data items, and the attribute values of the data items exist in the tables, figures and texts of the articles.
Each data item is about the description of a bioevents.
The participants need to find out some attributes' values of the bioevents, the attributes including sample id, profile name, locality, latitude, longitude, age and depth.	
We observed how participants used \KBS~for data extraction and verified the accuracy of the data they eventually extracted.
After the data extraction part, we had a 20-minute semi-structured interview with each participants about their experience with \KBS.
\begin{table*}
    \centering
    \resizebox{1.\linewidth}{!}{
    \begin{tabular}{c|l|c|c}
    \toprule
      Article ID &  Title & Pages & Data Item Number \\
    \midrule
      A1   & \makecell[l]{Age and synchronicity of planktonic foraminiferal bioevents across the Cenomanian–\\Turonian boundary interval (Late Cretaceous) \cite{falzoni2018age} } & 38 & 18\\
      \midrule
      A2   & \makecell[l]{An expanded Cretaceous−Tertiary transition in a pelagic setting of the Southern Alps\\ (central-western Tethys) \cite{fornaciari2007expanded}} & 34 & 18\\
      \midrule
      A3   & \makecell[l]{Palynology of the Cenomanian to lowermost Campanian (Upper Cretaceous) Chalk \\of the Trunch Borehole (Norfolk, UK) and a new dinoflagellate cyst bioevent\\ stratigraphy for NW Europe \cite{pearce2020palynology}} & 60 & 30\\
    \bottomrule
    \end{tabular}
    }
    \caption{The articles used for user study.}
    \label{tab:articles}
\end{table*}

\begin{figure}
    \centering
    \includegraphics[width=0.8\linewidth]{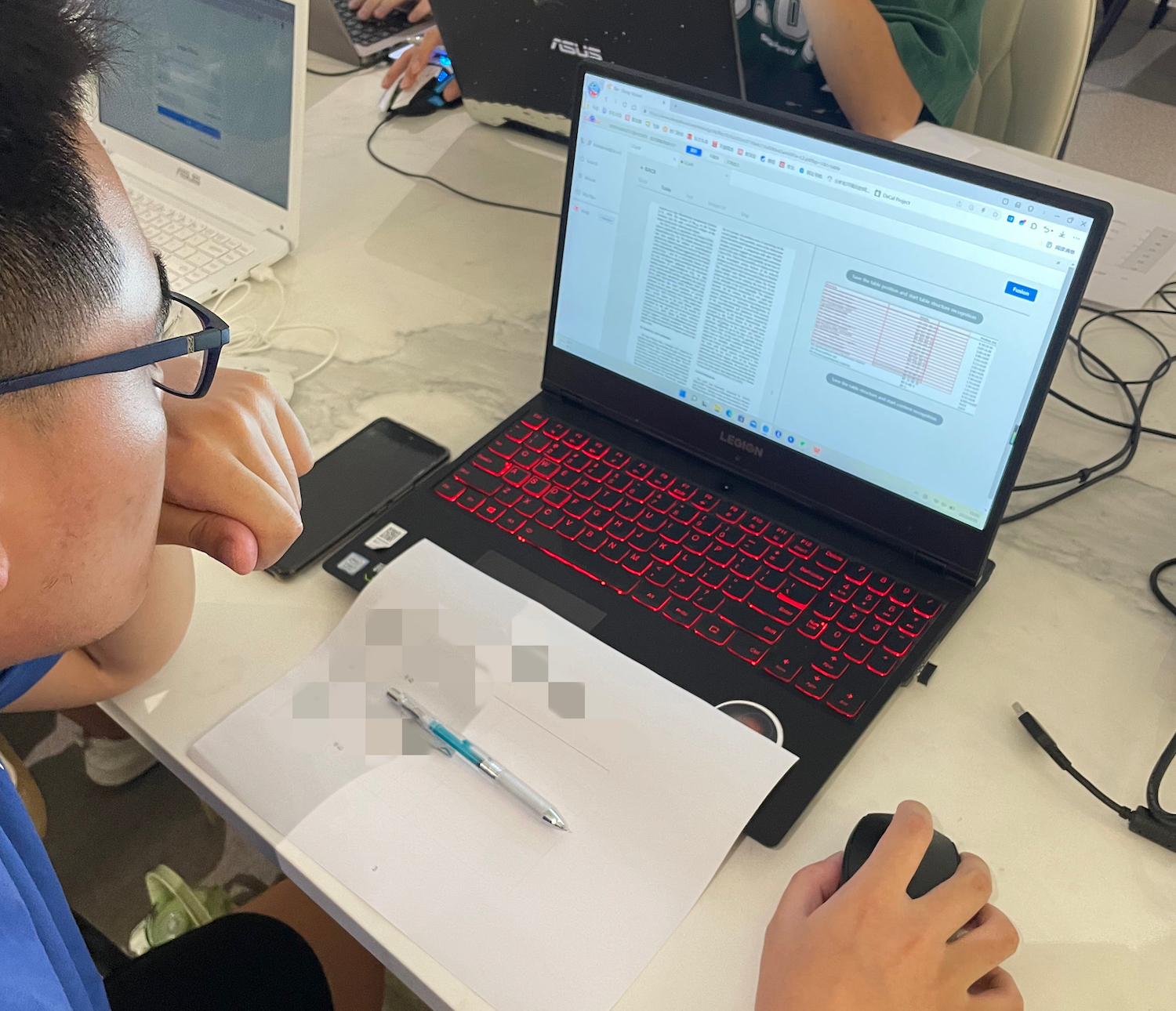}
    \caption{The photo of user study session, showing the participant using \KBS~to extract the data from a table in PDF. }
    \label{fig:evaluation-photo}
\end{figure}

\subsection{Results}
All 7 participants completed the data extraction task for the three articles. 
The average time to process A1 is about 17 minutes, A2 is 23 minutes, and A3 is 17 minutes.

We observed and recorded user behaviors during the experiment to understand how researchers use \KBS~to extract data from articles when building scientific knowledge bases.
We aim to explore the following questions:
(1) How did researchers locate the data they needed? 
(2) How did the researchers extract the data in the table? 
(3) How did researchers find data item attribute values in images and text?

After the data extraction section, we conducted a 20-minute semi-structured interview with each participant. 
In addition to understanding how they felt in the user study sessions, we also asked them about their current workflow with \KBS~and their willingness to use it and got their recommendations for the system.
All 7 participants agreed that \KBS~significantly improved their productivity in the data extraction task.

\subsubsection{\textbf{AI-in-the-loop can help improve the efficiency of locating the data.}}
During the locating data phase, participants performed a quick scan of the article, which was consistent across all participants. 
We found that they only focus on parts of the table when locating, which means that in the browsing process, they only look at whether the table in the article contains the relevant bioevent field. 
This situation is consistent with our findings in previous user research that researchers do not focus on the central idea of the article and the opinions expressed when doing data extraction--they only focus on the data itself.

All participants said that the function of table positioning could help them quickly find the table they needed to extract.
Both PC03 and PC05 mentioned that the table positioning function could make them more aware of the table and the data in it when browsing the article, and it is convenient to judge whether data extraction is required:

\textit{``\KBS~is much faster for the positioning of tables in the literature and the extraction of table data than manual work''.} -PC03

\textit{``Using \KBS~is faster than finding information by myself in the literature ''.} -PC05

The text extraction module in our system pre-extracts phrases containing place names and latitude and longitude after parsing the full text. 
We observed that all 6 users took advantage of this pre-extraction feature, and only one participant still used the segment-by-segment reading method for attribute value localization. 
In the data extraction from figures, the test article used in our user research includes map identification and location information extraction, and we observed that participants could successfully use this function to extract the critical point data.

In both pictures and text, the participants felt that \KBS~helped them find the data faster.
Several participants said that in the text function, the highlighting of pre-extracted words combined with their own experience allowed them to find the required data values in a large amount of text more quickly:

\textit{``When looking for some data in the text, targeted highlighting facilitates us to find key data values''.} -PC03

This situation also confirms that a general-purpose model in AI-in-the-loop practice can significantly improve efficiency in the extraction task and significantly reduce the cost compared to training a complex model in a human-in-the-loop fashion.

\subsubsection{\textbf{AI-in-the-loop can help improve the data accuracy by the design of multi-step human-AI collaboration pipeline.}}
\begin{figure}
    \centering
    \includegraphics[width=\linewidth]{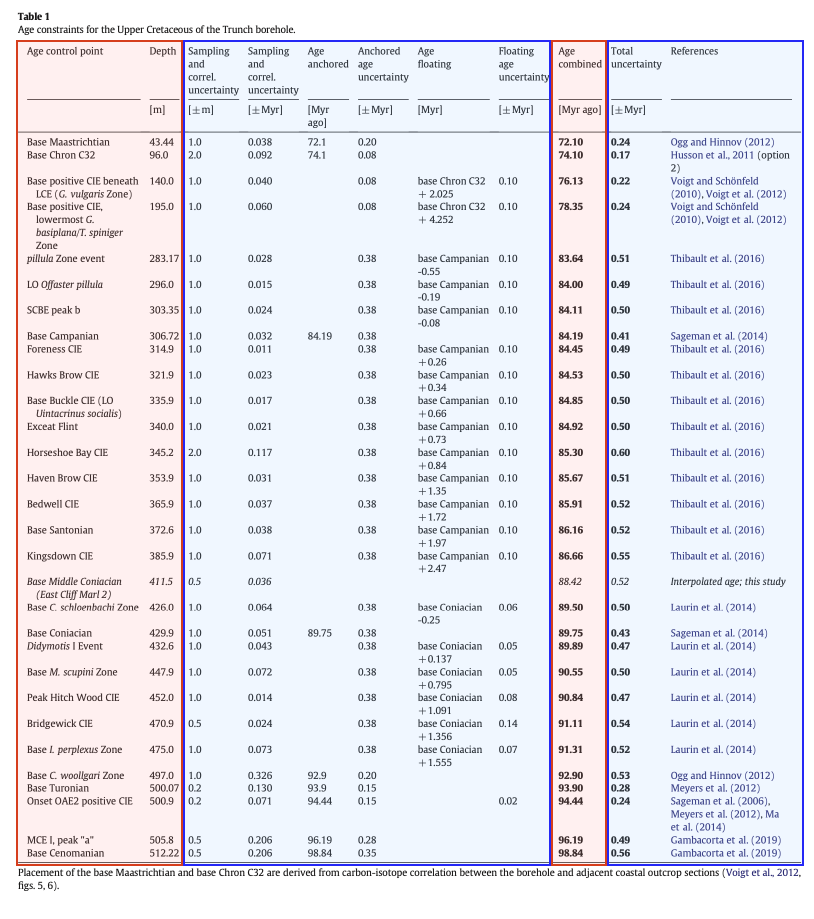}
    \caption{The target table from A3 \cite{pearce2020palynology}.The \textcolor{red}{red areas} are the data needed, and the \textcolor{blue}{blue areas} are the irrelevant data.}
    \label{fig:table-A3}
\end{figure}

In articles A1 and A2, the schema of the table is simple. 
All participants were entirely consistent with our design, including the entire table into the selection range, then using the complete extraction of the table for proofreading. 
However, in A3, the structure of the target table becomes complicated and contains some fields they do not care about (see Figure \ref{fig:table-A3}). 
Participants' behaviors varied during the extraction of this table. 
We observed that participants who were graduate students used a method of extracting only the required fields, while the lower-level undergraduate participants still fully extracted the entire table and then thought about which fields were needed. 
We believe this is related to differences in the domain knowledge of the participants. 
Participants with more domain knowledge were able to more quickly identify which fields in the form were needed, which influenced how they used them. 
Such a block extraction method is beyond our design assumptions. 
We think this is an interesting phenomenon, and there is room for expansion of the table extract functions.

Compared with the original manual extraction, all the participants think that using \KBS~greatly improves the speed of table processing.
PC07 believes that the advantages of \KBS~are very obvious when dealing with tables:

\textit{``In the past, we could only manually extract the values in the table one by one, but now using \KBS~I can directly get all the data at one time. Tables are merged directly by attribute name.''} -PC07

This advantage is also reflected in the processing of tables with large data volumes:

\textit{``For tables with large amounts of data, using \KBS~to extract data is much faster than manual extraction''.} -PC04

Moreover, PC06 mentioned that she had used another table extraction tool, but the accuracy was not acceptable:

\textit{``I have also used TABULA, which has fast recognition speed, but low recognition accuracy''.} -PC06

We further asked them for their views on step-by-step interactions.
All participants agreed that this method could effectively correct some minor errors. Furthermore, they agreed that the current model works very well in recognition of simple tables and only needs confirmation to complete the extraction, which is also consistent with our observation of their performance in the extraction process.
 PC05 made further observations:

\textit{``May be limited by the current recognition algorithm, some complex forms I need to adjust a little bit, I will find it a little difficult''.}

\subsection{Threats to Validity}
Our user research and system function design are both conducted in the field of geoscience.
The sample bias in our participants is a potential threat to the validity of the user study results.
When we select experimental materials, in order to avoid the situation that the function cannot be tested, we have made the data distribution in the article cover the figures, texts and tables.
In order to eliminate the influence of experimental material (selected articles) on user research, our participants came from the same research direction. 
This is also out of the professional consideration of the scientific knowledge base construction.
However, the singleness of the research field of the participants may still lead to some underrepresentation.

During the user study, we observed that the use behavior of researchers at different levels is also very different, which also has a certain impact on the efficiency and accuracy of actual use.
We think there will also be some interesting results from studying the differences in use of participants at different levels.

We plan to address these potential threats to validity in the future through (1) collect more usage data of different teams for analysis during the operation of the system, and compare the differences between different teams; (2) publish the system to more different disciplines to verify the effect of the system in more different research areas.

\section{Discussion}
The results from our user study and the practical application suggest that researchers can successfully collaborate with \KBS~to extract data from scientific literature and integrate the data to build a scientific knowledge base. 
In this section, we discuss the lessons we learned and the design implications of our work.

\subsection{AI-in-the-loop method in High Data Accuracy Task}

The rational use of human and model capabilities is the key to AI-in-the-loop systems. 
The ability of AI to provide helpful information in this process will be more important than the ability of AI itself to complete the task.

According to the results we observed in the user study, domain experts have effectively improved both efficiency and accuracy when using the AI-in-the-loop approach for data extraction.
Efficiency gains are due to AI's ability to retrieve and locate critical information in a large amount of information, which is reflected in the portion of the data needed to locate. We design models to quickly show domain experts in their simplest form where information is worth attention throughout the document.
The improvement in accuracy is due to the interaction pipeline between the model trained on the generic data and the human. 
Training on generic domain data saves the cost of collecting domain-specific annotated data. The domain experts' knowledge can be successfully used to proofread and confirm the final output quickly when the task can be basically completed since the model's output is designed as an adjustable process.
How the pipeline ensures the accuracy of the data will be discussed in detail in Section \ref{decision-making}.

\subsection{Decision Making in Human-AI Collaboration}\label{decision-making}

In an AI-in-the-loop system, humans frequently interact with AI. From the results of the user study, the  multi-step multi-modal human-AI collaboration pipeline used by \KBS~effectively realizes the cooperation between domain experts and AI.
The pipeline can first effectively help users understand the role of the model, allowing users to make decisions more intuitively as to whether they need to be modified.
Secondly,the design of the pipeline can avoid information overload, which helps the user focus more on the current decision.
Considering the interaction process between models and humans, preventing information overload and model overload, and making the interaction process as disassembled as possible to fit into the human decision-making process are two critical issues to be considered.
Since people have confirmed between each input and output of the model (only need to check and make minor adjustments), the model's input between each step is adjusted to the best state to ensure that the output results are more accurate.

Based on data accuracy requirements, we ensure that people have control over the data in the process of human-AI collaboration. In other words, we enable humans to always have the decision-making power to eliminate errors in the model output throughout the process of human-AI interaction.
During the whole process, the data will be continuously exchanged between the human and the model, and human decision-making can reasonably utilize the human's advantages in error judgment to eliminate the errors generated by the model.
Domain experts no longer need to rely too much on data scientists to adjust the model in the process of use, and they can directly adjust the data itself to interact with the model.

Considering the perspective of domain experts on AI, this pipeline is also better than the human-in-the-loop method. According to the results of formative research, domain experts have a large gap between the mental model of AI and the actual conceptual model. Compared with the end-to-end approach of separating humans from AI in the human-in-the-loop method, interacting with the AI model through this pipeline can also make domain experts more agree with the improvement brought by AI in this work.

\subsection{Future Works}

\subsubsection{Opportunity for a better Human-AI Collaboration}

Based on the human-AI interaction pipeline we discussed in Section \ref{decision-making}, we believe that AI-in-the-loop can explore a better human-AI collaboration practice that can combine the advantages of human-in-the-loop.
The AI-in-the-loop approach has been proven in our study to provide researchers with higher efficiency and data accuracy. 
We think the next step is to verify that this method can help improve the model. 

In the field of artificial intelligence, human-in-the-loop research focuses on how to further improve the model's performance and does not focus on the experience of domain experts and crowd workers. 
In the actual deployment applications, the model ultimately serves the research and work of domain experts. 
We found in formative research and PD sessions that domain experts did not get a good experience as the ultimate user or beneficiary, which may be the existence of human-in-the-loop methods in the actual deployment of a problem. 

We believe that exploring the AI-in-the-loop practice in the deployment stage, and on this basis, exploring the advantages of combining the human-in-the-loop method, that is, using human feedback to promote the improvement of the model, will be a beneficial method for better results.
Based on the step-by-step interaction feature, \KBS~can collect feedback data from users at every step, which can be considered accurately labeled data for the task done by the current model.
Most directly, these user feedback data can be used to retrain the model to improve the model accuracy, which is often used by some human-in-the-loop methods.
Furthermore, since the AI model deployed by AI-in-the-loop in \KBS~is a model for basic general tasks, it can be used to fine-tune the model in such a specific field of tasks, making the model more suitable for specific fields tasks and further improving the performance of the model.

We found that the AI-in-the-loop approach adopted by \KBS~has the opportunity to increase AI-assisted efficiency in human systems while collecting accurate human feedback data (which can be considered accurate labeling data).
This method can not only complete the task of domain experts collecting scientific data to build a scientific knowledge base, but also collect data that can be used for model training, further improving the efficiency and performance of the human-AI team in completing tasks.
Exploring how feedback data can play a role in model training and iteration will help test this idea.

\subsubsection{Supporting Data Processing in Big-data Driven Discovery}
We plan to expand \KBS~to more natural science disciplines' big data research.
The current version of \KBS~is primarily based on user research related to earth sciences, but clearly, this is a general job.
We hope to explore the need for scientific knowledge base construction in more disciplines and further improve the extraction of scientific data from the literature.
In addition, we also found in the PD session that the data in the pictures in the scientific literature is also very rich, but the current vision-based related methods have not achieved better extraction results. We believe that this is also a direction of future development. Participants in the PD Session believe that the extraction and restoration of data graphs can mine the measurement data of some older documents, contributing to document digitization and data rescue.
This direction also applies to other experimental disciplines. 
In our preliminary research, we also found that most of the experimental-based disciplines contain a large number of data charts in the literature, which are used to depict the original measurement data.
This requirement is also confirmed in some previous studies \cite{10.1145/3025453.3025957,luo2021chartocr}.

\section{Conclusion}
In this paper, we practice the framework of AI-in-the-loop in building scientific knowledge bases. 
We propose \KBS, an AI-assisted system that uses neural network models to support data extraction from literature in PDF format. 
We summarize the workflow and design the system based on formative study and participatory design sessions involving domain experts. 
The system supports the extraction of metadata, data in tables, data in figures and texts of literature. 
Our practise and user study prove that the AI-in-the-loop method can reduce the burden of experts while ensuring data accuracy. 
Moreover, based on our design in the multi-step multi-modal human-AI collaboration pipeline, we believe that feedback from experts can be collected effectively, which is expected to combine some practical experience of the human-in-the-loop method with \KBS~in the future for improving the model performance in the experts’ workflow. 
We hope that our work can further push for better human-AI collaboration on the task of scientific knowledge bases construction.

\begin{acks}

We owe a particular debt of gratitude to the scientists from the Deep-time Digital Earth program who all contributed enormously valuable feedback. 
We also thank Jia Guo, Tao Shi, Yifei Shen, Le Zhou, Qi Li, Zhixin Guo, Mingxuan Yan, Mingze Li, Jingyao Tang, Han Liu and Shengling Zhu for their support for our system development.
This work is supported by National Natural Science Foundation of China (No.42050105, No.62106141) and Shanghai Sailing Program (21YF1421900).
This work is a part of the Deep-time Digital Earth (DDE) Big Science Program.

\end{acks}

\bibliographystyle{ACM-Reference-Format}
\bibliography{sample-base}

\appendix

\section{Questionnaire used in Formative Study}\label{questinnaire}

\begin{table}[H]
    \centering
    \begin{tabular}{ll}
    \toprule
     QID & Questions\\
    \midrule
    \multicolumn{2}{l}{\textbf{Part 1: Basic Information}}\\
    Q1 & Gender\\
    Q2 & Career Position\\
    Q3 & Research Field in Geoscience\\
    \midrule 
    \multicolumn{2}{l}{\textbf{Part 2: Task- related Information}}\\
    Q4 & The progress of Data extraction\\
    Q5 & Research Team size\\
    Q6 & Team composition\\
    Q7 & Tools using in data extraction\\ 
    Q8 & Data source used\\
    Q9 & Data source format and proportion\\
    Q10 & Method to process PDF files\\
    Q11 & Distribution and proportion of data in the file\\
    Q12 & Number of database fields\\
    Q13 & The number of documents needed to build the knowledge base\\
    Q14 & Personal participation in data collection\\
    Q15 & Current Workflow of data extraction\\
    Q16 & Time of Single PDF processing\\
    Q17 & Estimate the time required for the entire collection\\
    \midrule 
    \multicolumn{2}{l}{\textbf{Part 3: Understanding of computer technology}}\\
    Q18 & Understanding of programming\\
    Q19 & Kinds of tasks be accomplished by programming\\
    Q20 & Understanding of artificial intelligence\\
    Q21 & Data set preparation of artificial intelligence task\\
    Q22 & Understanding of data labeling\\
    Q23 & Kinds of tasks served by data labeling\\
  \bottomrule
    \end{tabular}
    \caption{Questions in the questionnaire.}
    \label{tab:Questionnaire}
\end{table}

\section{System Implementation Details}\label{data_extraction_model}
\KBS's front-end interactive single-page web application is built in Vue.js and is hosted by Nginx.
Python and the FastAPI framework are used to implement \KBS's back-end API service. The asyn- chronous coding approach enables higher concurrency with less resource usage, allowing it to accommodate more users concurrently.
To store documents and extracted data, we use a master-slave backup MySQL database, which offers comprehensive security and efficient reading and writing.
Considering user system security, we only store and bcrypt hashed passwords to ensure that users' plaintext passwords will not be stored and leaked.
Furthermore, the HTTPS protocol is applied to the whole system of \KBS~to ensure the security in network communication.

\subsection{Document Management \& Retrieval}\label{documentmangement}
All the literature uploaded into \KBS~are all automatically parsed with Grobid \cite{GROBID} and Science Parse \cite{tkaczyk2018machine}.
The meta information of papers (e.g., Title, Author List, Abstract, Venue, and Year) is extracted and indexed with Elasticsearch.
Then all the fields could be utilized for searching and retrieving the documents.
Moreover, to better browse and manage literature, the document list could be filtered with the principal user as well as the import user and sorted by title, import time, and the latest update time.
To go further, each user could get ``My File List'' containing only documents taken charge by him and ``Recent File List'' containing his recent viewed documents, which allows users to obtain the documents most important to them and simply continue their respective workflows.
\begin{figure}[H]
    \centering
    \includegraphics[width=\linewidth]{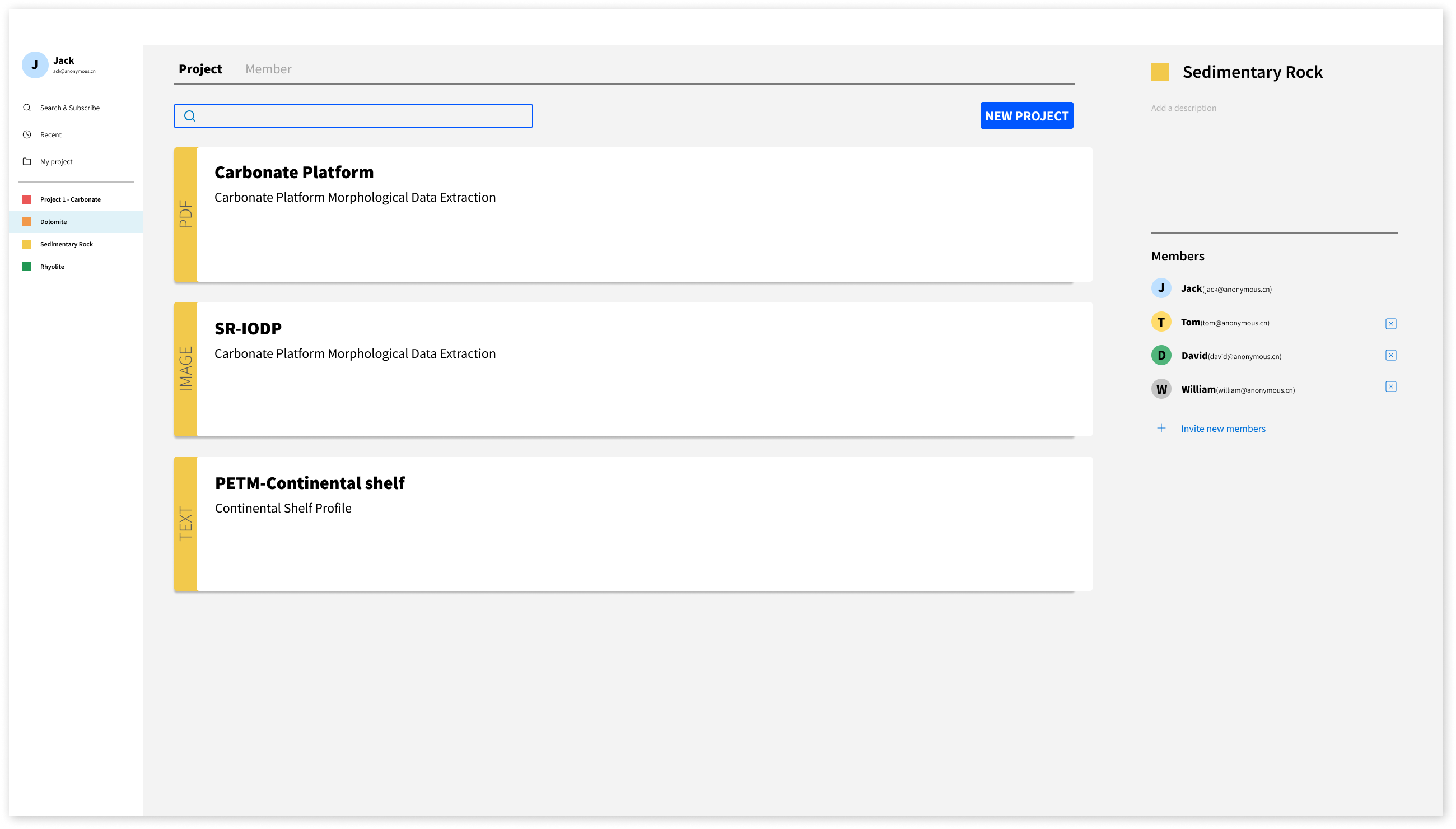}
    \caption{The interface of project list page in \KBS.}
    \label{fig:projectlistl}
\end{figure}
\begin{figure}
    \centering
    \includegraphics[width=\linewidth]{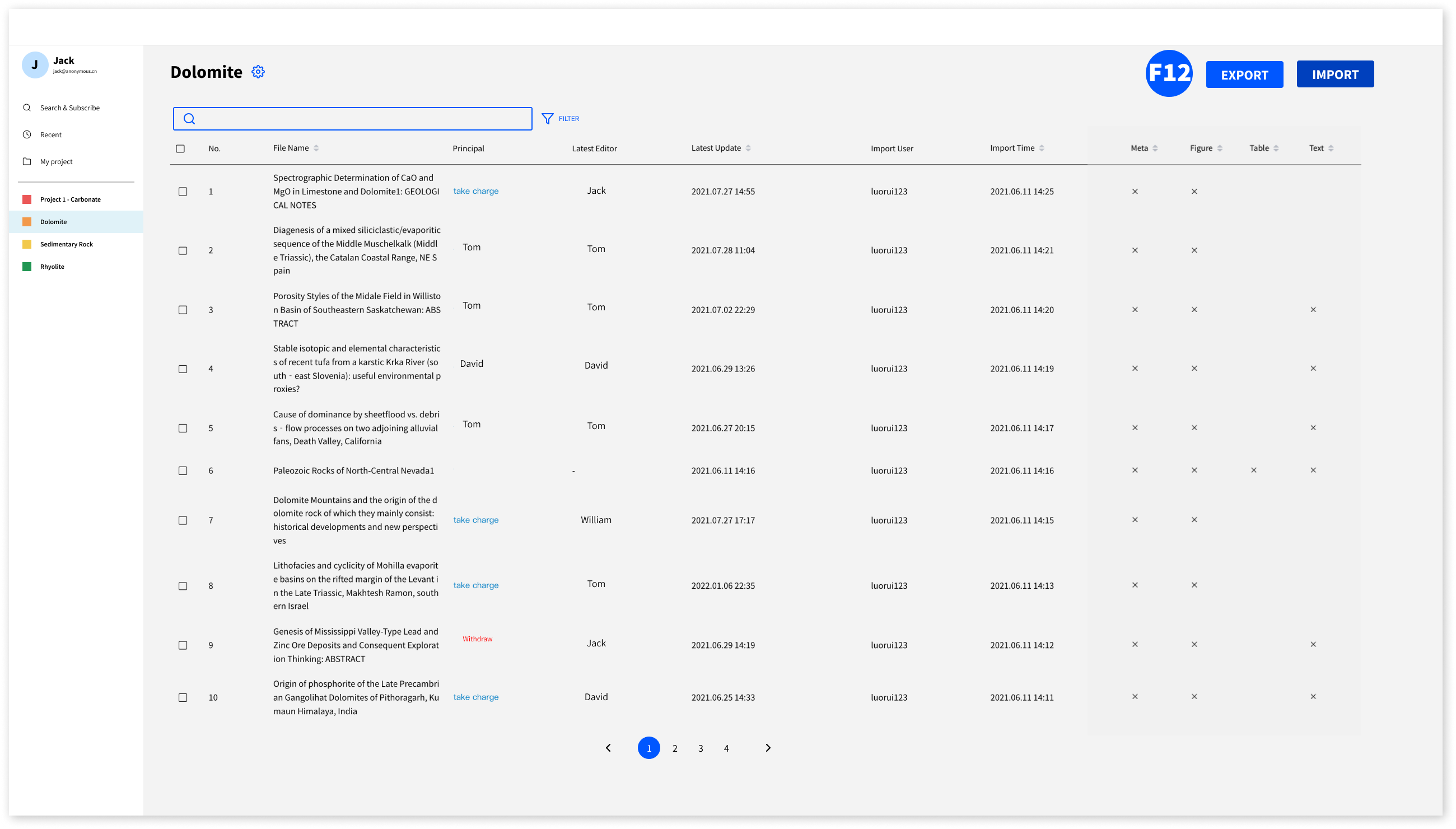}
    \caption{The interface of file list page in \KBS.}
    \label{fig:filelist}
\end{figure}

\subsection{Data Extraction}\label{extraction-implementation}
Generally \KBS~extract data from these parts from literature:
\begin{itemize}
    \item \textbf{Meta Information Extraction} For each uploaded document, \KBS~uses multiple parsing tools (e.g., Grobid, Science Parse, and PdfFigures 2.0) to independently extract its meta information and mix all the information with a voting mechanism.
    \item \textbf{Table Extraction:} First, \KBS~uses an object detection model Detectron2 \cite{wu2019detectron2} trained on TableBank \cite{li2019tablebank}, a benchmark dataset for table detection, to detect the region of tables. Then for each table, a series of rules are adopted to locate each cell within it. Once users confirm the cell structure of a table, Tesseract \cite{10.5555/1288165.1288167} will be applied to detect the text in each cell and establish the final digitalized table.
    \item \textbf{Text Extraction:} To extract academic entities from papers with the format of PDF, \KBS~first utilizes PDFFigures 2.0 \cite{clark2016pdffigures} to parse each text section from the original files. Then some rules and the natural language processing library spaCy \cite{spacy2} are adopted to automatically extract entities of different types from the parsed text sections.
    \item \textbf{Location Extraction in Figure:} Users can box the region of any map they care about. Then \KBS~will detect the longitude and latitude labeled at the margin of the map and determine the coordinate range of the entire map. Then if users click any location on the map, the exact coordinates of the location will be automatically calculated and recorded.
\end{itemize}

\end{document}